\documentclass[aps,prb,twocolumn,amssymb,floats,epsfig]{revtex4}
\usepackage{epsfig,psfrag,subfigure}
\usepackage{graphicx,psfrag,subfigure}
\usepackage{color}
\usepackage{braket}
\usepackage{amssymb,amsbsy,amsmath,bbold}

\newcommand\beq{\begin{equation}}
\newcommand\eeq{\end{equation}}
\newcommand\bea{\begin{eqnarray}}
\newcommand\eea{\end{eqnarray}}
\newcommand\al{\alpha}

\newcommand\ga{\gamma}
\newcommand\de{\delta}

\newcommand\De{\Delta}
\newcommand\si{\sigma}

\newcommand\om{\omega}
\newcommand\ta{\theta}
\newcommand\dg{\dagger}
\newcommand\pa{\partial}
\newcommand\non{\nonumber}
\newcommand\noi{\noindent}

\newcommand\bi{\bibitem}
\newcommand{\floor}[1]{\lfloor #1 \rfloor}

\begin{document}

\title{Critical phase boundaries of static and periodically kicked long-range 
Kitaev chain} 

\author{Utso Bhattacharya$^1$, Somnath Maity$^1$, Amit Dutta$^1$ and 
Diptiman Sen$^2$}
\affiliation{$^1$Department of Physics, Indian Institute of Technology, 
Kanpur 208016, India \\
$^2$Centre for High Energy Physics, Indian Institute of Science, Bengaluru
560012, India}


\begin{abstract}
We study the static and dynamical properties of a long-range Kitaev 
chain, i.e., a $p$-wave superconducting chain in which the superconducting 
pairing decays algebraically as $1/l^{\al}$, where $l$ is the distance 
between the two sites and $\al$ is a positive constant. Considering very 
large system sizes, we show that when $\al >1$, the system is topologically 
equivalent to the short-range Kitaev chain with massless Majorana modes at 
the ends of the system; on the contrary, for $\al <1$, there exist symmetry 
protected massive Dirac end modes. We further study the dynamical phase 
boundary of the model when periodic $\de$-function kicks are applied to the 
chemical potential; we specially focus on the case $\al >1$ and analyze the 
corresponding Floquet quasienergies. Interestingly, we find that new 
topologically protected massless end modes are generated at the quasienergy 
$\pi/T$ (where $T$ is the time period of driving) in addition to the end 
modes at zero energies which exist in the static case. By varying the 
frequency of kicking, we can produce topological phase transitions between 
different dynamical phases. Finally, we propose some bulk topological 
invariants which correctly predict the number of massless end modes at 
quasienergies equal to 0 and $\pi/T$ for a periodically kicked system with 
$\al > 1$.
\end{abstract}

\maketitle

\section{Introduction}
\label{sec1}

Topological phases of quantum condensed matter systems have been extensively 
studied for the past several years~\cite{hasan,qi,fidkowski}. Typically, 
these are phases in which the system is gapped and insulating in the bulk 
but has gapless modes at the boundaries which can contribute to transport
at low temperatures. Further, the number of species of gapless boundary 
modes is given by a topological invariant whose nature depends on the spatial 
dimensionality of the system and the symmetries that it possesses, such as 
chiral symmetry, particle-hole symmetry and time-reversal symmetry. The 
significance of a topological invariant is that it does not change if the 
system is perturbed as long as the bulk modes remain gapped and the symmetry 
of the system remain unaffected by the perturbation. Two- and three-dimensional
topological insulators, quantum Hall systems, and one-dimensional wires with 
$p$-wave superconductivity provide some examples of topological phases.

More recently, there has been considerable interest in exploring 
topological features associated with periodically driven closed quantum 
systems ~\cite{kita1,lind1,jiang,gu,kita2,lind2,trif,russo,basti1,liu,tong,
cayssol,rudner,basti2,tomka,gomez,dora,katan,kundu,basti3,schmidt,reynoso,
wu,thakurathi}. Boundary modes and topological invariants have been 
studied in this context~\cite{kita1,lind1,jiang,trif,liu,tong,cayssol,rudner,
kundu}. A photonic topological insulator has been demonstrated experimentally 
where a two-dimensional lattice of helical waveguides has been shown to 
exhibit topologically protected end modes~\cite{recht}. However, the 
existence of topological invariants and the relation between them and the 
number of Majorana modes at the boundary seems to be unclear, particularly if 
the driving frequency is small~\cite{tong}. Further, the Majorana boundary 
modes are of two types (corresponding to eigenvalues of the Floquet operator 
being equal to $+1$ or $-1$). In a recent study of a periodically driven 
one-dimensional short-range Kitaev (SRK) chain~\cite{kitaev}, it has been shown 
that the numbers of these two types of modes can be obtained from a 
topological invariant~\cite{thakurathi}. In some variations of the Kitaev 
chain, end modes corresponding to Floquet eigenvalues {\it not} equal to 
$\pm 1$ are found~\cite{saha}; it is not known if there are topological 
invariants which can describe such modes.

In this paper, we will study in detail one-dimensional $p$-wave 
superconductors with long-range pairings. We will first revisit
its static phase diagram and then investigate the dynamical generation of end 
modes under periodic kicking of the chemical potential. We will elaborate on
how the presence of long-range pairings modify the properties of both the 
static and the dynamical systems, especially in comparison with the 
corresponding short-range systems~\cite{thakurathi}.

Let us recall that topological superconductors are expected to host massless 
Majorana modes which obey non-Abelian statistics and can therefore serve as 
essential components of a topological quantum computer. The most important
feature of such topological systems is the existence of robust gapless 
boundary modes which are immune to disorder. There have been a number
of experiments recently which have found signatures of Majorana modes
in one-dimensional or quasi-one-dimensional systems~\cite{mourik}.

Considering the relevance and immense interest in realizing topological 
superconductors, the Kitaev chain which is a paradigmatic model exhibiting 
topological superconducting phases has been studied in the presence of 
long-range superconducting 
pairings~\cite{vodola14,vodola16,viyuela14,viyuela16,huang14}
to determine which specific properties of a model are necessary for such phases
to occur. In this model, the superconducting pairing is taken to decay as 
$1/l^{\al}$, where $l$ is the distance between the two sites. It has been
reported that there are three different topological phases depending upon the 
power-law exponent $\al$ which characterizes the pairing. For $\al>3/2$, it 
has been established that the model is topologically equivalent to the 
SRK chain with nearest-neighbor pairing terms. However, for 
$\al<1$, the emergence of a new topological phase has been noted which hosts 
massive Dirac fermionic end modes that lie within the bulk energy gap and 
are topologically robust~\cite{viyuela16}. Most intriguing is the presence of 
a crossover region that interpolates smoothly between the $\al<1$ and the 
$\al>3/2$ regions. To better understand the behavior of the crossover region, 
we have performed exact diagonalization for large system sizes in this paper 
to establish that the crossover region $1<\al<3/2$ only exists as 
a consequence of the system size being taken to be finite in earlier papers. 
We exhibit that in the thermodynamic limit of the system size $L \to \infty$, 
this intermediate region is also topologically equivalent to the SRK chain 
characterized by a well-defined topological invariant which is a winding 
number and is always an integer. Therefore there are only two different 
regions: a short-range model for $\al > 1$ and a long-range model for $\al < 1$.

To attain further control over topological phases of matter, one sometimes 
considers periodically driven systems governed by the Floquet theorem. The 
periodic driving offers such an immense control over topological phases that 
it shows considerable promise as a viable approach towards creating topological 
materials with high tunability and engineering new non-equilibrium topological 
phases which are absent in their static counterparts. For instance, Majorana 
end modes can be generated in a one-dimensional $p$-wave superconducting 
system, by varying the chemical potential through some periodic $\de$-function 
kicks with time~\cite{thakurathi}. Motivated by these considerations,
we investigate in this paper how the topological properties of the Kitaev 
chain get modified in the presence of both periodic kicking and long-range 
superconducting pairing with a power-law decay and whether some exotic
topological phases emerge in this driven system. 

The outline of this paper is as follows. In Sec.~\ref{sec2} we introduce the 
system of interest and review some of its properties. The system we consider 
is a tight-binding model of spinless electrons with a long-range $p$-wave 
superconducting pairing and a chemical potential $\mu$; this is called the
long-range Kitaev (LRK) chain. In Sec.~\ref{sec3}, we discuss the energy 
spectrum, the phase boundaries and the topological and non-topological phases 
that this model possesses. In this process we establish the phase 
diagram of the LRK chain and we illustrate that it consists of two regions 
based on the value of $\al$: (i) for $\al>1$, the LRK chain is effectively 
short-ranged and has topological phases which are characterized by a 
topological invariant (winding number) given by $\nu = 1$ and host massless 
(zero energy) Majorana modes at the ends of a long system for $-1 < \mu < 1$, 
while (ii) for $\al<1$, the long-range pairing gives rise to topological 
phases which have winding number given by $\nu = 1/2$ and host massive Dirac 
modes at the ends for $\mu > -1$~\cite{viyuela16}. 
On the other hand, the system has topologically trivial 
phases which have no end modes for $|\mu|>1$ if $\al>1$ and $\mu < -1$ if 
$\al<1$; these phases are characterized by a winding number equal to $0$ and 
$-1/2$ respectively. In Sec.~\ref{sec4}, we discuss our analytical method of 
studying the Floquet evolution and derive the Floquet Hamiltonian when the 
the chemical potential is driven by periodic $\de$-function kicks which obey 
all the symmetries of the static Hamiltonian~\cite{stock}. In Sec.~\ref{sec5}, 
we study the quasienergy spectrum versus the chemical potential for various 
values of the system parameters and show that new topological phases emerge. 
We also establish the critical gap closing points for certain ranges of $\al$ 
before discussing the ubiquitous features generated by the periodic kicking.
In Sec.~\ref{sec6}, we study the ranges of parameters in which massless
Majorana and massive end modes appear at the ends of an open system. We then 
use the Floquet operator for a system with periodic boundary conditions to
define some topological invariants and show that they correctly predict the 
numbers of end modes with Floquet eigenvalues equal to $+1$ and $-1$ for the 
$\al>1$ region. We find that end modes can either appear or disappear as the 
driving frequency is varied. Finally we summarize our main results and 
point out some directions for future work in Sec.~\ref{sec7}.

\section{Kitaev chain with long-range superconducting pairing}
\label{sec2}

We consider a model of spinless fermions on a one-dimensional chain with 
long-range $p$-wave superconducting pairings; this is known as the LRK chain. 
For a system with $L$ sites and periodic boundary conditions, the Hamiltonian 
has the form~\cite{vodola14}
\beq \label{eq_ham}
\begin{split}
H & = ~\sum_{n=1}^{L} \Big\{ \ga \left(c_{n+1}^\dg c_{n} ~+~ c_{n}^\dg c_{n+1}
\right) ~-~ \mu ~(2c_n^\dg c_n-1) \\
& ~~~~~~~~~~+ ~\sum_{l=1}^{L-1} ~\frac{\De}{d_l^\al} ~(c_{n+l}^\dg c_n^\dg ~+~
c_n c_{n+l})\Big\}, \end{split} \eeq
where $\ga>0$ is the hopping amplitude, $\mu$ is the chemical potential,
$\De$ is the superconducting pairing amplitude, and $c_n$ ($c^\dg_n$) are 
fermionic annihilation (creation) operators defined at site $n$ of the 
chain. (We will assume that both $\ga$ and $\De$ are real and positive; $\mu$ 
can be positive or negative). The superconducting pairing is a function of the 
distance $d_l= \text{Min}[l,L-l]$ between the two sites $n$ and $n+l$; it is
long-range since it decays as a power-law with an exponent $\al > 0$. Although 
the Hamiltonian does not conserve the total fermionic number, the parity 
operator (total fermionic number modulo two) commutes with the Hamiltonian 
and is conserved.

For both periodic and open boundary conditions, the Hamiltonian in 
Eq.~\eqref{eq_ham} can be written in terms of Majorana operators. 
The Majorana operators are defined as follows,
\bea a_{2n-1} &=& c_n + c^{\dg}_n, \non \\
a_{2n} &=& i \left(c_n - c^{\dg}_n\right). \eea
It is easy to verify that these are Hermitian operators satisfying 
$\left\{a_m,a_n\right\} = 2\de_{mn}$. Using these operators, we can rewrite 
the Hamiltonian (Eq.~\eqref{eq_ham}) with open boundary conditions as
\begin{widetext}
\beq \label{eq_hamm}
H ~=~ \sum_{n=1}^{L-1}{\left[\frac{i\ga}{2}\left(a_{2n}a_{2n+1}-a_{2n-1}a_{2n+2}
\right) ~-~ \frac{i\De}{2} ~\sum_{l=1}^{L-n}\frac{1}{l^\al}\left(a_{2n}
a_{2n+2l-1} +a_{2n-1}a_{2n+2l}\right)\right] } ~+~ i\mu ~\sum_{n=1}^{L} 
a_{2n-1}a_{2n}. \eeq \\
\end{widetext}

We first discuss the well established extreme short-range limit 
($\al \to \infty$)~\cite{kitaev} of Eq.~\eqref{eq_hamm}. There are two 
phases present in this limit: a trivial phase for $|\mu|>1$, and a topological 
phase for $-1<\mu<1$. In the topological phase, there is one Majorana zero 
mode (MZM) at each end of an infinitely long open chain. (If the chain
is not infinitely long, the modes at the two ends hybridize and their
energies shifts from away zero; then the modes do not have a 
Majorana character). The two MZMs can together be written in terms 
of an ordinary fermion operator (for which $c \ne c^\dg$); the 
corresponding zero energy state may be occupied or unoccupied ($c^\dg c = 1$ 
or 0). This gives rise to two possible ground states which are degenerate for
an infinitely long chain. Hence the ground state of the SRK chain is two-fold 
degenerate, but the two states have different parities depending on whether 
the zero energy state is occupied or not. For instance, if the bulk has 
an even number of occupied states, the two possible parity 
sectors are given by (i) states where the MZM is unoccupied resulting in 
even fermion parity, and (ii) states where the MZM is occupied giving rise
to odd fermion parity. We will show below that the LRK chain remains 
effectively short-ranged as long as the decay exponent $\al > 1$; this
is in contrast to the results obtained in Ref.~\onlinecite{viyuela16}.

Without loss of generality, we will set $\ga=-\De=1$ in the rest of this 
paper. If we assume periodic boundary conditions, the translational symmetry 
of the system enables us to implement a Fourier transformation and rewrite the 
Hamiltonian in the Nambu spinor basis given by $\psi_k=(c_k, c^\dg_{-k})^T$. 
In the thermodynamic limt $L \to \infty$, the momentum $k$ becomes a continuous 
variable lying in the interval $[-\pi,\pi]$, and we 
get~\cite{vodola14,vodola16,viyuela16}, 
\beq H ~=~ \int_{-\pi}^\pi ~\frac{dk}{2\pi} ~\Psi_k^\dg H_k \Psi_k, \eeq
where
\beq H_k ~=~ - ~f_\al(k) ~\si^y ~+ \left(\cos{k} - \mu\right)~ \si^z, 
\label{hamk} \eeq
and 
\beq f_\al(k) ~=~ \sum_{l=1}^{\infty} ~\frac{\sin{(kl)}}{l^\al}. 
\label{falk} \eeq
This is related to the polylogarithm function $Li_\al (z) \equiv 
\sum_{l=1}^\infty z^l/l^\al$ as
\beq f_\al(k) ~=~ - \frac{i}{2} ~\left(Li_\al(e^{ik}) - Li_\al(e^{−ik}) 
\right). \eeq
The eigenvalues of the Hamiltonian in Eq.~\eqref{hamk} are 
\beq \label{eq_spectrum}
E_k^\pm ~=~ \pm ~\sqrt{\left( \cos{k} - \mu \right)^2 ~+~ \left( f_\al(k)
\right)^2}. \eeq
To simplify the notation, we now rewrite the Hamiltonian in terms of a
unit vector $\hat{n}_k$ so that $H_k$ takes the form
\beq \label{eq_hk}
H_k ~=~ \De_k ~\hat{n}_k\cdot \vec{\si}, \eeq
where $\vec{\si} = (\si^x, \si^y, \si^z)$ are the Pauli matrices, and 
\bea \De_k &=& |E_k^\pm|, \non \\
\hat{n}_k &=& \frac{1}{\De_k} ~\left(0, - f_\al(k), \cos{k} - \mu\right). \eea

It is worth noting that the LRK chain lies in the BDI symmetry class of 
topological insulators and superconductors~\cite{schnyder08,kitaev09}, and 
is particle-hole, time-reversal, and chiral symmetric. These symmetries 
restrict the movement of the vector $\hat{n}_k$ from a sphere $S_2$ in
three dimensions to a circle $S_1$ in the $y-z$ plane; this leads to a
mapping from the Hamiltonians $H_k$ in the Brillouin zone $k \in S_1$ 
to the winding vector $\hat{n}_k \in S_1$. This mapping, for a short-ranged 
interacting system, yields a $Z$-valued topological invariant called the 
winding number $\nu$ which is the angle subtended by $\hat{n}_k$
when the momentum $k$ goes from $-\pi$ to $\pi$. We find that
\beq \label{eq_winding}
\nu ~=~ \frac{1}{2\pi} ~\oint ~dk ~\frac{\pa_k n^y_k}{n^z_k}. \eeq
Alternatively, the same winding number can be obtained by considering an 
adiabatic transport of the system from a certain crystal momentum through
a reciprocal lattice vector. After making the Hamiltonian in Eq.~\eqref{eq_hk}
off-diagonal by a unitary rotation, the eigenstate in the lower band,
$|g_k\rangle = (u_k, v_k)^T$, then picks up a Berry/Zak 
phase~\cite{berry84,aharonov87,zak89} defined as 
\beq \phi_Z~=~ i ~\oint ~dk ~\langle g_k|\pa_k|g_k\rangle. \label{zak} \eeq
This is generally quantized in integer multiples of $\pi$ and has a one-to-one
correspondence with the winding number defined in Eq.~\eqref{eq_winding}.

\section{Phase diagram}
\label{sec3}

\begin{figure*}[]
\centering
\subfigure[]{\includegraphics[width=.44\textwidth,height=7.7cm]{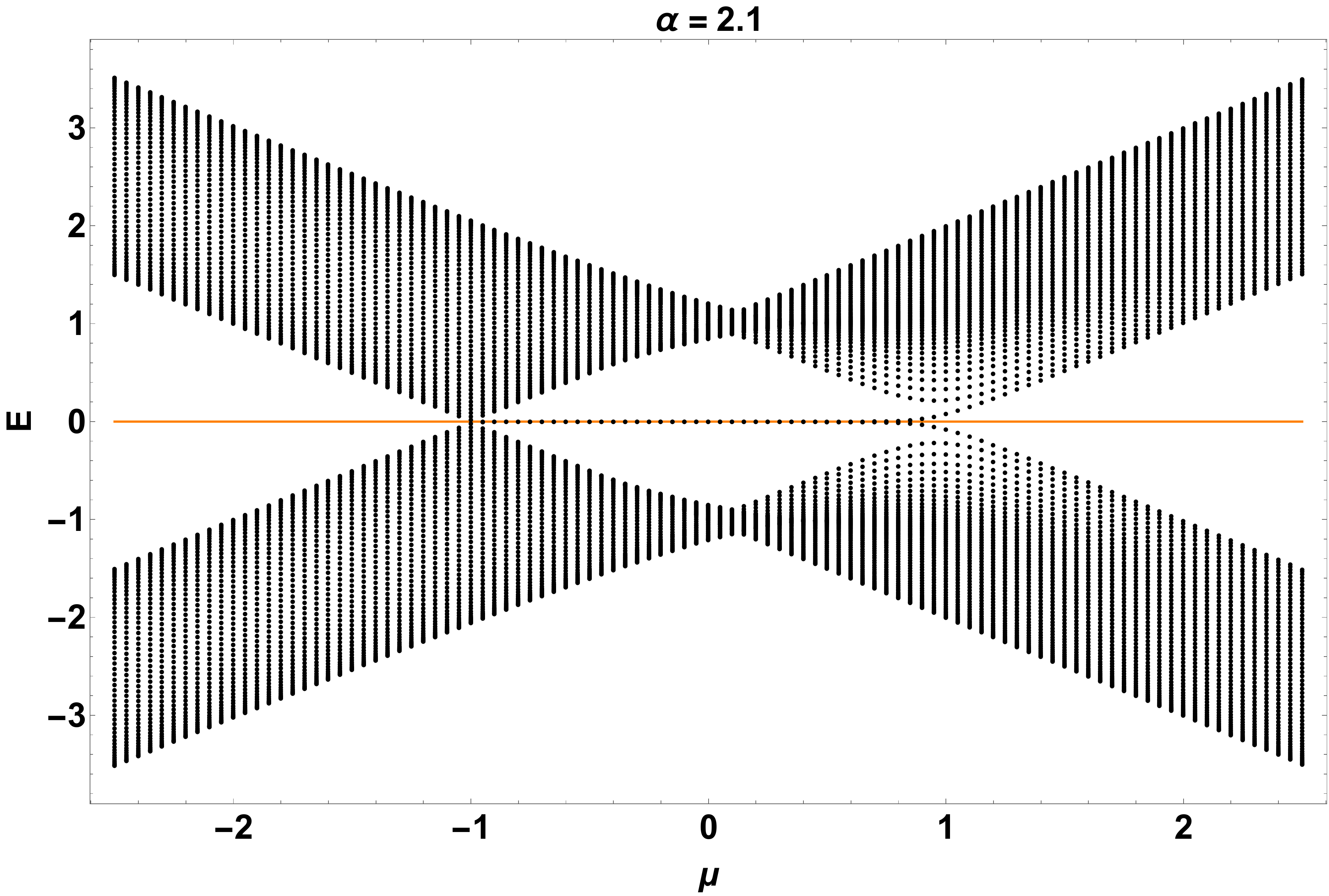}
\label{UE1a}}
\hfill
\quad
\subfigure[]{\includegraphics[width=.45\textwidth,height=8cm]{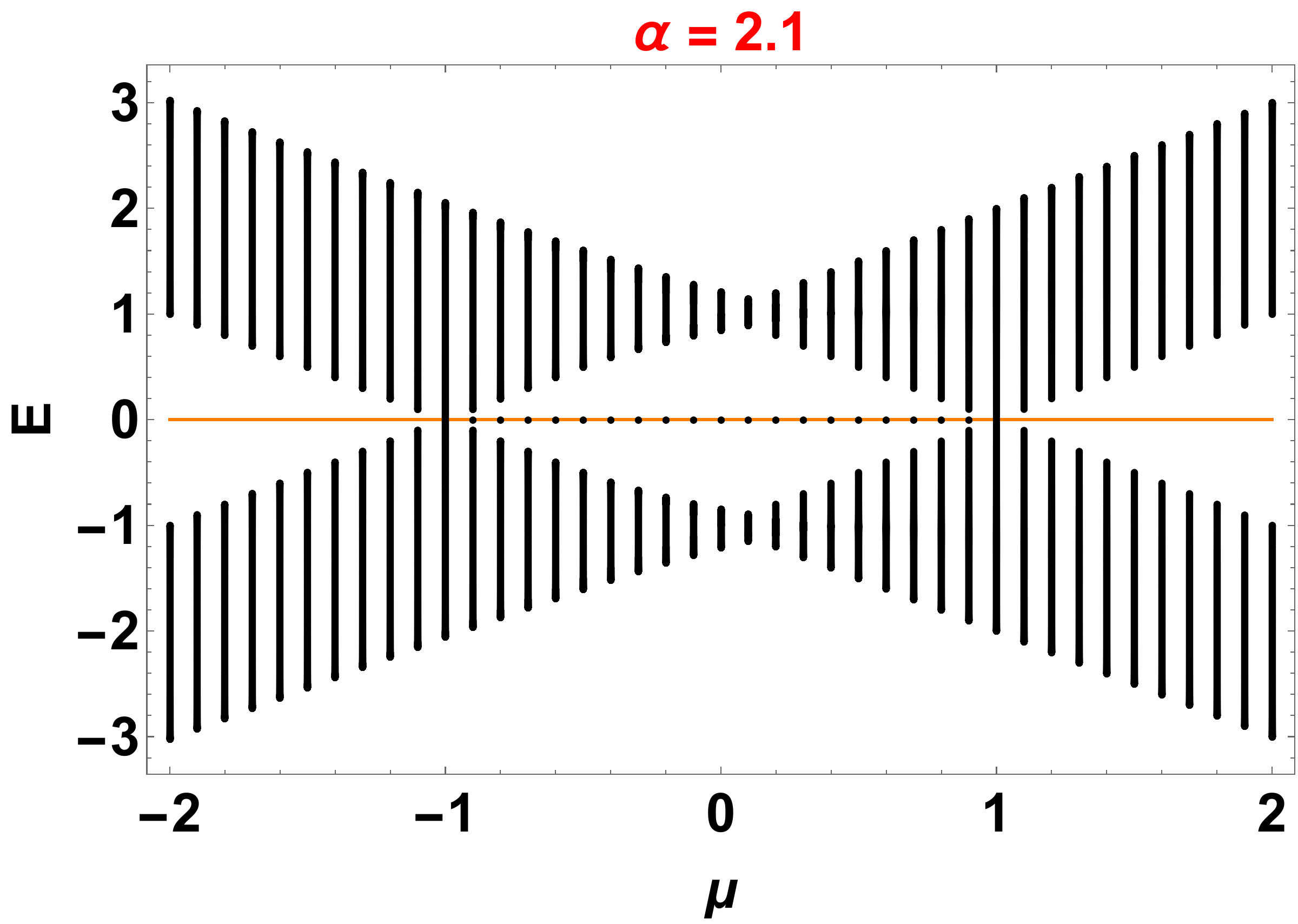}
\label{UE1b}}
\caption{(a) Energy spectrum $E$ of the LRK chain with open boundary 
conditions plotted versus the chemical potential $\mu$ for $\al = 2.1$ and 
system size $L = 60$. We see that there is a slight energy gap at $\mu = 1$.
(b) Energy spectrum $E$ for the same value of $\al = 2.1$ but with 
a much larger system size $L = 1000$. On increasing the system size we 
see that the slight energy gap at $\mu = 1$ has disappeared, up to numerical 
precision. The energy spectrum for this value of $\al$ is topologically 
equivalent to the SRK chain as it hosts zero energy Majorana modes for 
$-1<\mu<1$ and is non-topological otherwise.} \label{UE1} \end{figure*}

In order to determine the topological phase diagram of this system, we
find the critical lines of the Hamiltonian by considering the thermodynamic 
limit $L \to \infty$. 
The critical lines are obtained by studying the zeros and divergences of 
the function $f_\al(k)$ in Eq.~\eqref{falk} which encodes all the information 
about the long-range pairing. It is evident that when $\al>1$, $f_\al(k_c)=0$ 
for the critical momenta $k_c=0$ and $\pi$, but when $\al \leq 1$, 
$f_\al(k_c)=0$ only for $k_c=\pi$. For $\al < 1$, $f_\al (k)$ diverges at
zero momentum i.e., $f_\al(k)\to\infty$ as $k\to 0$. This leads to a divergence
of the energy in Eq.~\eqref{eq_spectrum} which scales with momentum as 
$E_k^\pm \sim k^{-(1-\al)}$. We thus see that for $\al > 1$, the energy goes 
to zero for both $k_c = 0,\pi$; hence both the lines $\mu=\pm 1$ are critical. 
On the contrary, due to the divergence of the energy at $k=0$ for $\al\leq 1$, 
only the line $\mu=-1$ is critical. Therefore, according to the behavior of 
$f_\al (k)$ at $k=0$, the existence of two different topological phases 
depending on the exponent $\al$ are obtained as follows. \\

\noi (a) The $\al > 1$ region is equivalent to the topological phase of the 
SRK chain~\cite{kitaev} as we will show below. The $|\mu| > 1$ phase is 
topologically trivial and is marked by the absence of MZMs. On the other hand, 
MZMs are present for $-1 < \mu < 1$. The presence of a phase discontinuity at 
$k=0$ in the ground state eigenvector $|g_k\rangle$ and the function 
$f_\al (k)$ being non-divergent yields the invariant $\nu = \phi_Z /\pi =
1$~\cite{viyuela16}. More generally, the winding number $\nu = 1$ for 
$-1<\mu<1$ and is zero otherwise, in the entire region $\al>1$.

To probe the behavior of the topological end modes in the 
$\al>1$ region, we will consider the two sub-regions $\al>3/2$ and 
$1 < \al < 3/2$. We first look at the region $\al>3/2$.
It is straightforward to numerically ascertain through the 
plots of the energy spectrum versus $\mu$ with open boundary conditions, 
that the region $\al>3/2$ is effectively short-ranged. In the plot 
of the energy spectrum $E$ against $\mu$ in Fig.~\ref{UE1a}, we notice that 
there is a slight gap in the energy spectrum around $\mu=1$; however this is 
an artefact of the relative smallness of the system size $L = 60$ chosen. 
Upon increasing the system size to $L=1000$ in (Fig.~\ref{UE1b}), we see that 
the slight energy gap at $\mu = 1$ disappears up to numerical precision. This 
establishes that the energy spectrum for $\al > 3/2$ is topologically 
equivalent to the SRK chain as it hosts MZMs for $-1<\mu<1$ and is 
non-topological otherwise. 

Next, we will look at the region $1 < \al < 3/2$. We will show that this
also hosts MZMs for $-1<\mu < 1$. However this range of $\al$ deserves special 
attention. This is because the group velocity $v_g = \pa_k 
E^{\pm}_k$ scales as $v_g \sim k^{-(3-2\al)}$ as $k\to 0$ and diverges for 
all $\mu$, whereas such a divergence of $v_g$ is absent when $\al > 3/2$.

\begin{figure*}[]
\centering
\subfigure[]{%
\includegraphics[width=.5\textwidth,height =8cm]{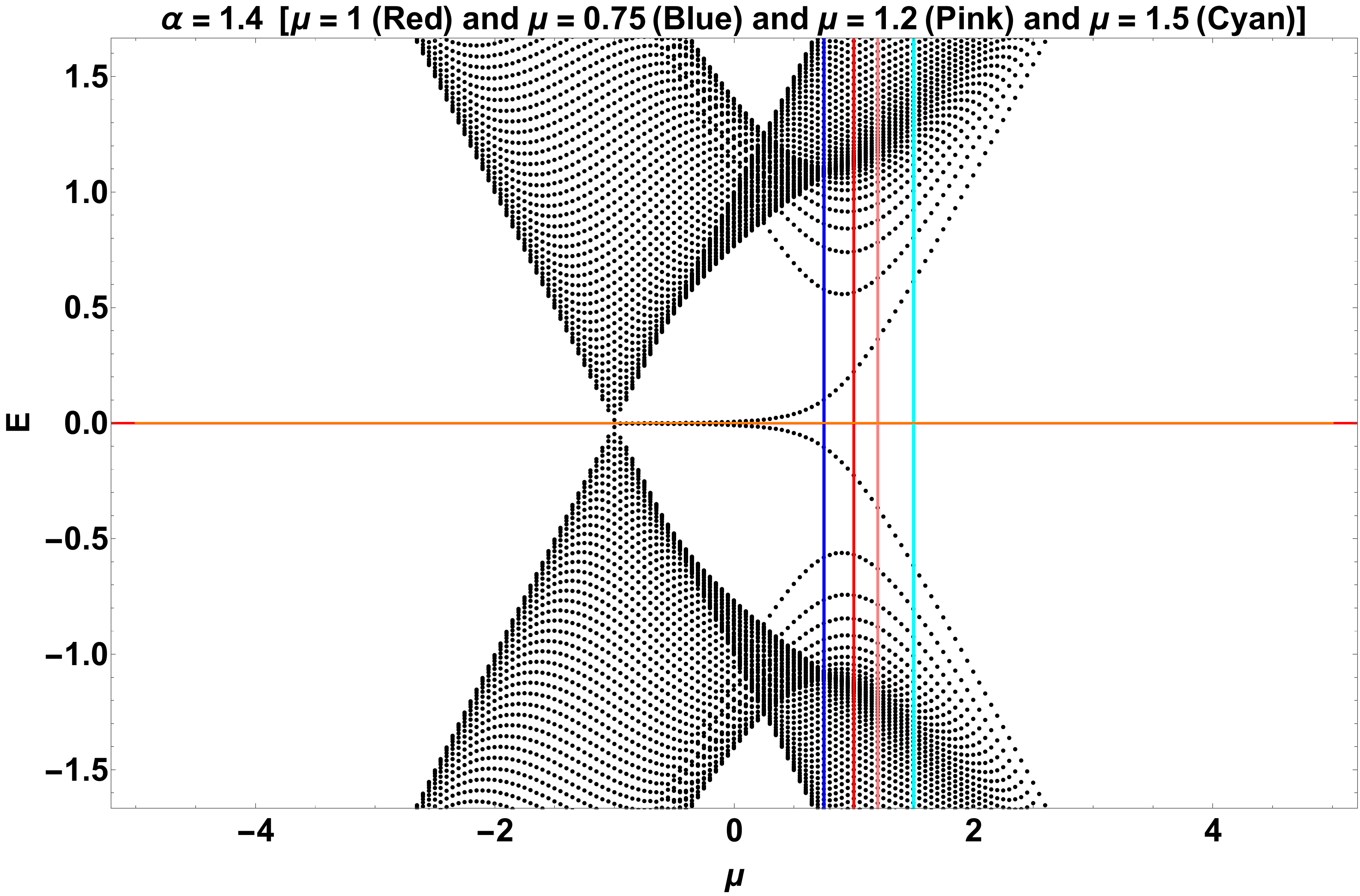}
\label{UE2L60}}
\hfill
\quad
\subfigure[]{%
\includegraphics[width=.4\textwidth,height=8cm]{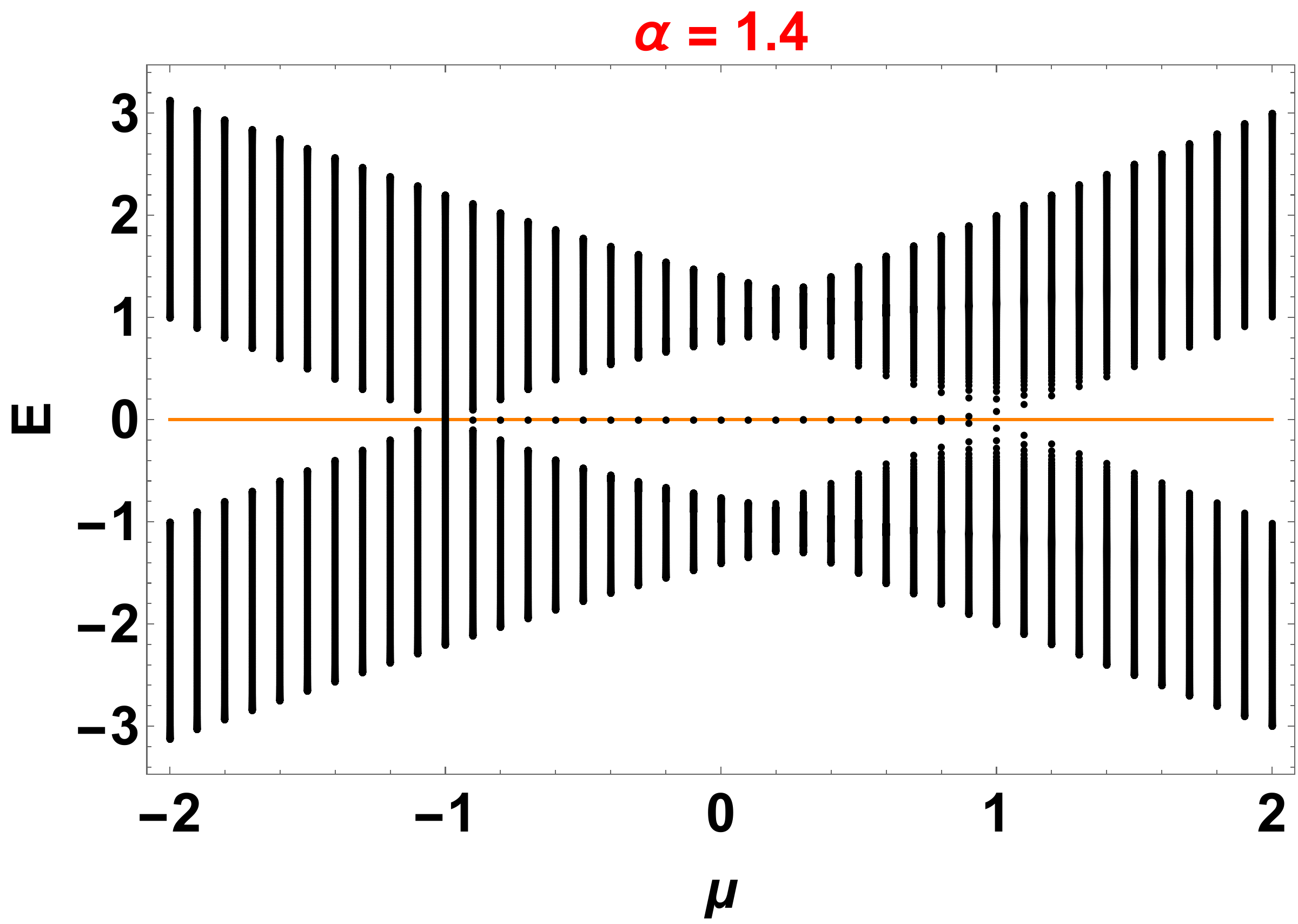}
\label{UE2L1000}}
\centering
\subfigure[]{%
\includegraphics[width=.48\textwidth,height=8cm]{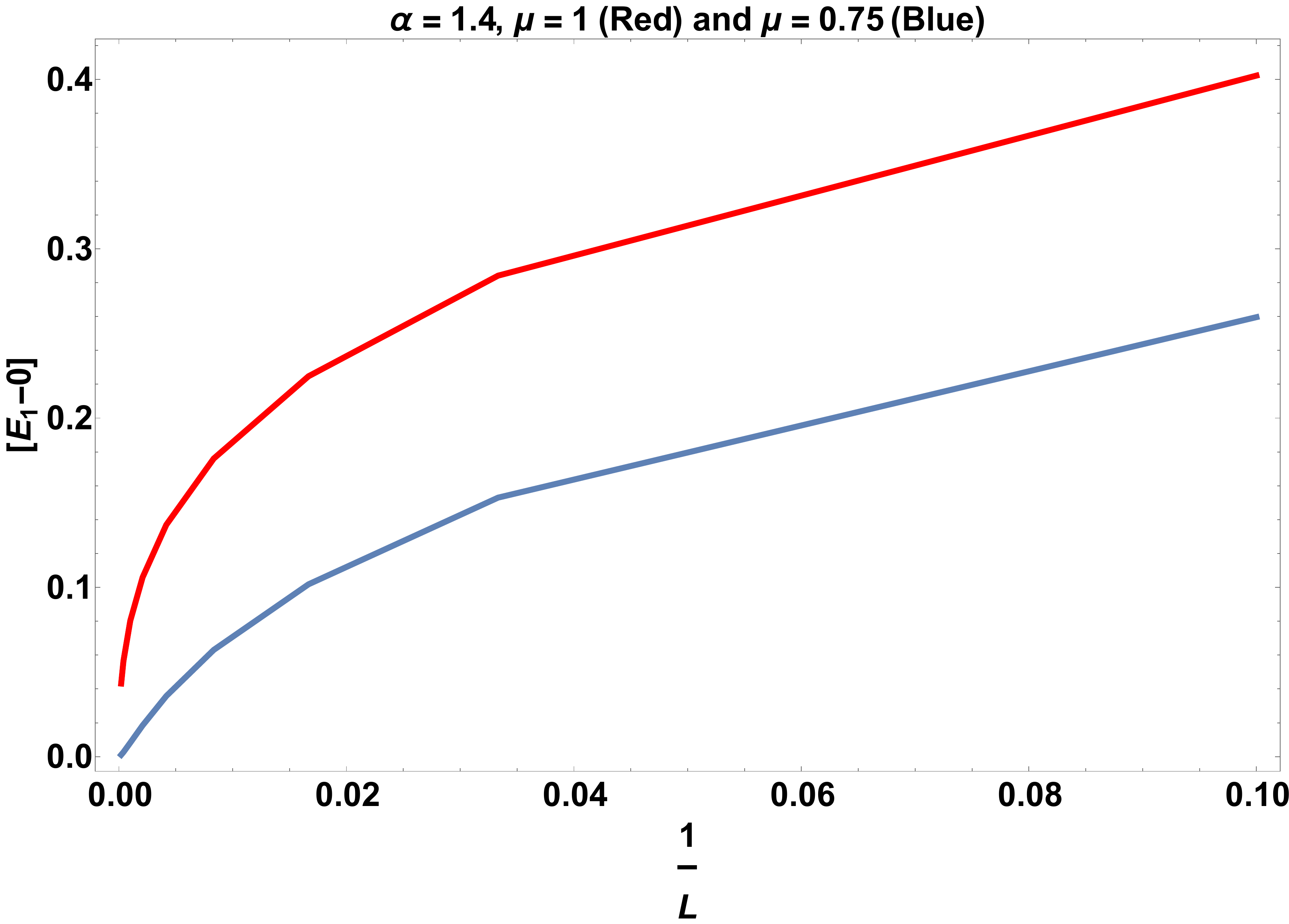}
\label{UE2ml1}}
\hfill
\quad
\subfigure[]{%
\includegraphics[width=.48\textwidth,height=8cm]{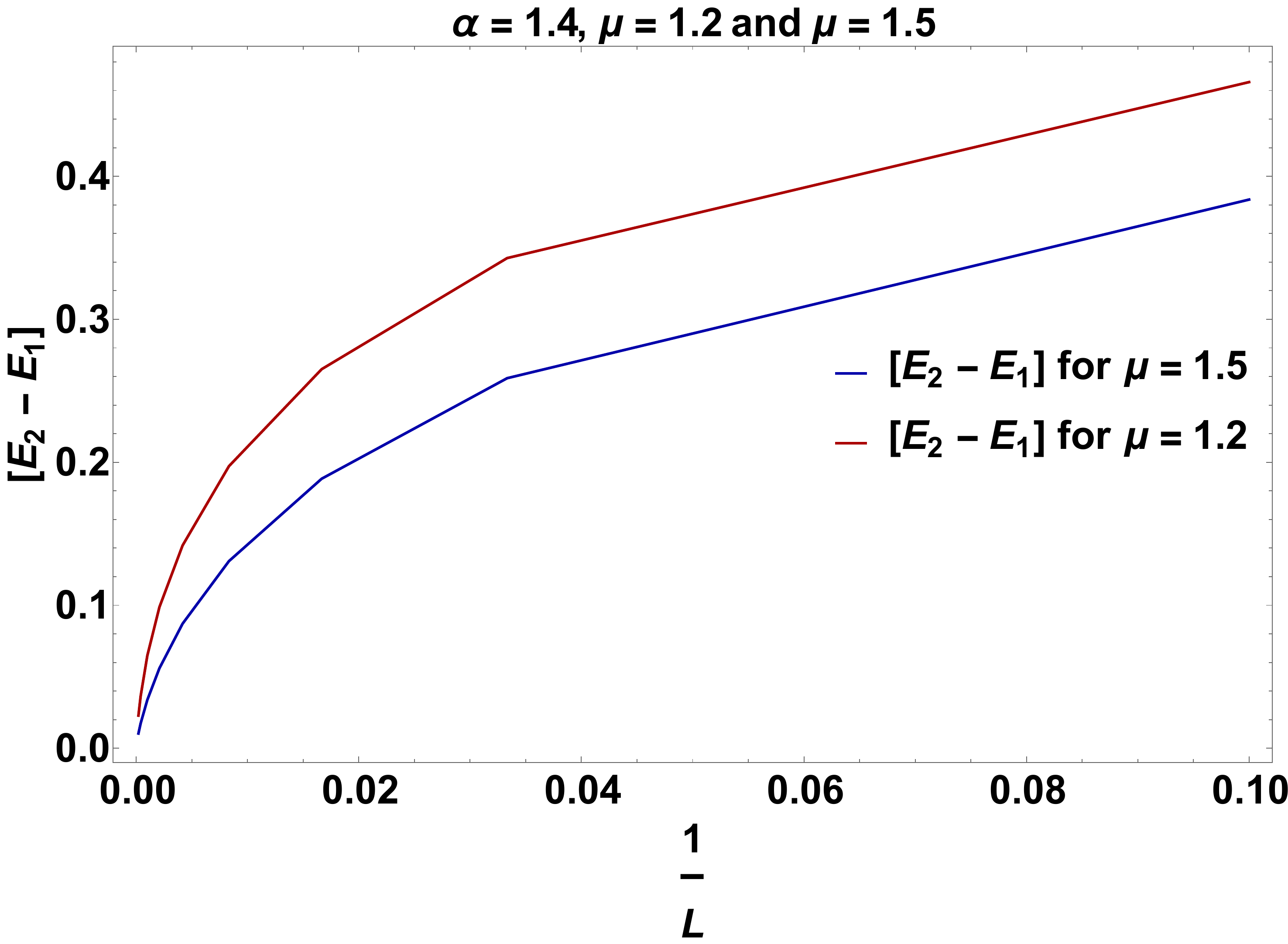}
\label{UE2mg1}}
\caption{(a) Energy spectrum $E$ of the LRK chain with open boundary 
conditions plotted versus the chemical potential $\mu$ for $\al = 1.4$ and for
a small system size $L = 60$. Three topological phases seem to exist depending
on the value of $\mu$: trivial with no end modes for $\mu < -1$, MZMs 
for $-1 < \mu \lesssim 0.5$, and MDMs
for $\mu \gtrsim 0.5$. The blue, red, pink, cyan vertical lines lie at 
$\mu = 0.75, 1, 1.2, 1.5$ respectively. The spectrum apparently exhibits a 
crossover-like behavior for $\mu \gtrsim 0.5$. (b) Energy spectrum $E$ of the 
LRK chain for the same value of $\al = 1.4$ but with a much larger system 
size $L = 1000$. The apparent crossover region observed for $L=60$ (Fig.~(a)) 
has almost completely vanished for $\mu>-1$, and there are almost 
no MDMs for any $\mu$. MZMs are however present for 
$-1<\mu<1$ establishing that for $\al =1.4$, the energy 
spectrum topologically resembles that of a SRK chain. (c) Deviation of mass 
gap (lowest positive eigenvalue $E_1$) from zero plotted versus the 
inverse of system size $1/L$ for two values of $\mu$ within the range 
$-1<\mu<1$. Although for a small system size $L=10$, there is a finite 
mass gap, the gap reduces with increasing system size and approaches zero for 
$L\to 5000$. The energy spectrum for $\mu = 0.75$ (blue) and $\mu = 1$ (red) 
would thereby host MZMs in the $L \to \infty$ limit. (d) The difference 
between the lowest two positive eigenvalues, $E_2$ and $E_1$, plotted versus 
$1/L$ to show how the MDMs merge with the bulk spectrum in 
the thermodynamic limit for two values of $\mu>1$, $\mu = 1.2$ (red) and 
$\mu = 1.5$ (blue). Although $E_2 - E_1$ is finite for a small system size 
$L=10$, the difference decreases with increasing system size as $L$ approaches 
$5000$. We therefore conclude that there would be no MDMs for
$L \to \infty$.} \label{fig_UE2} \end{figure*}

In Fig.~\ref{UE2L60}, we have plotted the energy spectrum $E$ of the LRK 
chain with open boundary conditions versus the chemical potential $\mu$ for 
$\al = 1.4$ for a small system size $L = 60$. This figure 
seems to imply that there are three topologically distinct phases depending 
on the value of $\mu$: trivial with no end modes for $\mu < 1$ and $\mu > 1$, 
and massless Majorana end modes and massive Dirac end modes (MDMs) which 
lie within the bulk gap for $-1 < \mu < 1$. The Dirac-like end modes
with energy slightly away from zero are visible near $\mu = 1$.
Thus the energy spectrum found observed for a small system size like $L=60$ 
appears to exhibit a crossover-like behavior around $\mu = 1$ where the 
energies of some end modes move away from zero and merge with the bulk bands. 
However this view of the spectrum is misleading. Since the group velocity 
$v_g$ diverges at $k = 0$ as $k^{-(3 - 2\al)}$, the energy spectrum becomes 
highly dispersive; for a finite size system in which $k$ is quantized in 
units of $2\pi/L$, the difference between the energies for two neighboring
values of $k$ diverges near $k=0$. Therefore, it takes an ever increasing 
system size $L$ to realize that a crossover region does not really exist in 
the thermodynamic limit $L \to \infty$. In Fig.~\ref{UE2L1000}, we show the 
energy spectrum $E$ of the LRK chain with open boundary conditions for 
$\al= 1.4$ and a much larger system size with $L = 1000$. It is now clear 
that the apparent crossover behavior observed for $L=60$ near $\mu =1$ has 
vanished; there are no MDMs at any value of $\mu$. However,
MZMs are present for $-1<\mu<1$ establishing that for $1< \al <3/2$, the 
energy spectrum topologically resembles that of a SRK chain.

To establish more conclusively the effectively short-ranged behavior 
of the model with $1< \al <3/2$ in the thermodynamic limit, we show in 
Fig.~\ref{UE2ml1} that for $-1<\mu<1$, the end modes become MZMs. 
The masslessness of the end modes in the region 
$-1 < \mu <1$ is verified by observing the variation of the mass gap 
(lowest positive eigenvalue $E_1$ minus zero) with the inverse of 
system size, $1/L$, for two values of $\mu$ lying in this region.
Although for a small system size $L=10$, there is a finite mass gap, the 
mass gap decreases with increasing system size and approaches zero for 
$L\to 5000$. The energy spectrum for $|\mu|<1$ would therefore host MZMs in 
the $L \to \infty$ limit. For $\mu > 1$, the difference between 
the lowest positive eigenvalue $E_1$ from the second lowest positive 
eigenvalue $E_2$ has been plotted versus $1/L$ in Fig.~\ref{UE2mg1}.
Although $E_2 - E_1$ is finite for $L=10$, the gap decreases with 
increasing system size as $L$ approaches the thermodynamic limit, showing 
that the MDMs merge with the bulk spectrum in the thermodynamic limit. Hence 
MDMs which would live near the ends of the system and would be separated from 
the bulk modes do not survive in the limit $L \to \infty$. Thus there are no 
end modes for $\mu > 1$. This makes the short-ranged nature evident for the 
entire range of values of $\al>1$. \\

\begin{figure*}[]
\centering
\subfigure[]{%
\includegraphics[width=.45\textwidth,height=8cm]{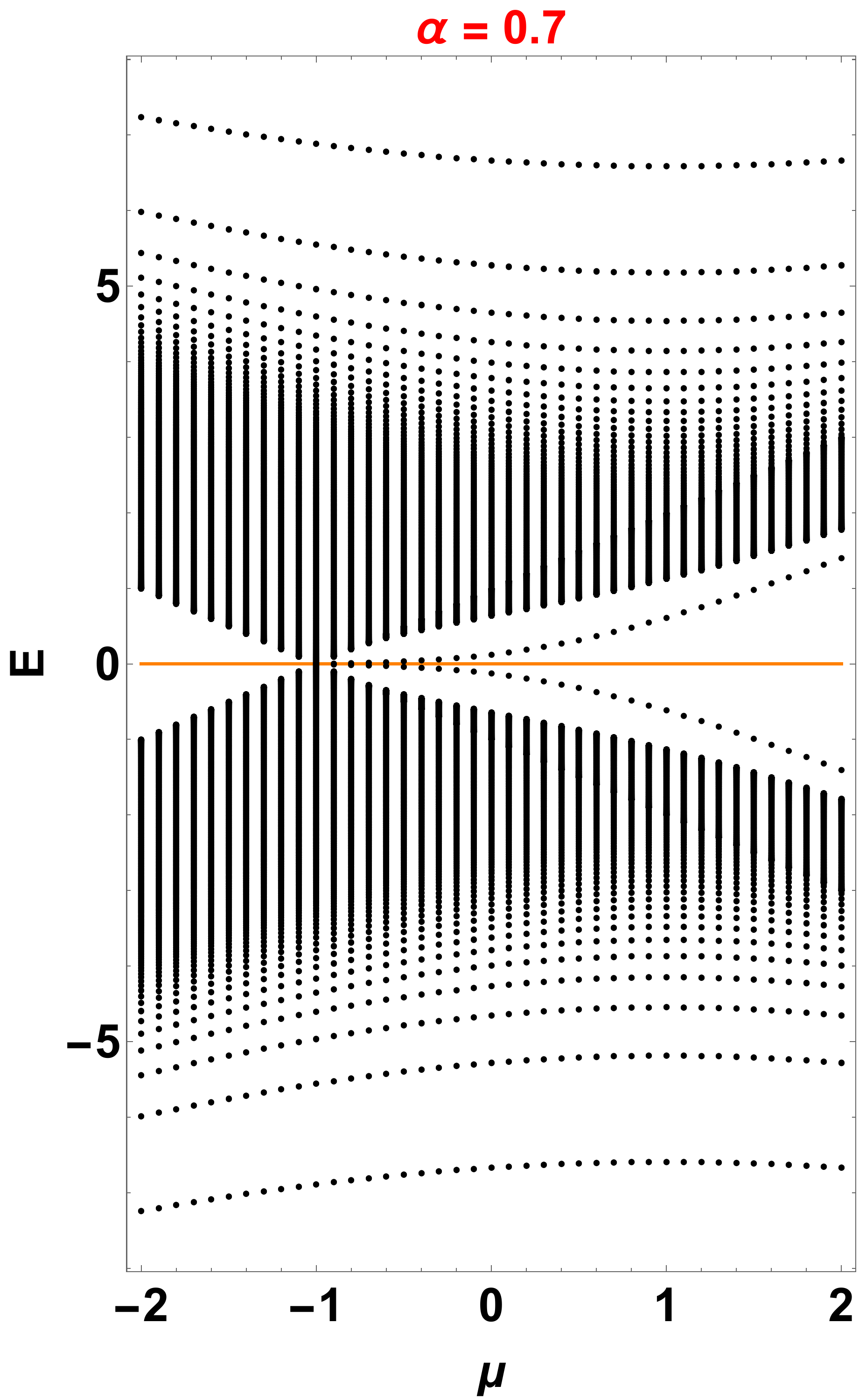}
\label{UEL}}
\hfill
\quad
\subfigure[]{%
\includegraphics[width=.45\textwidth,height=8cm]{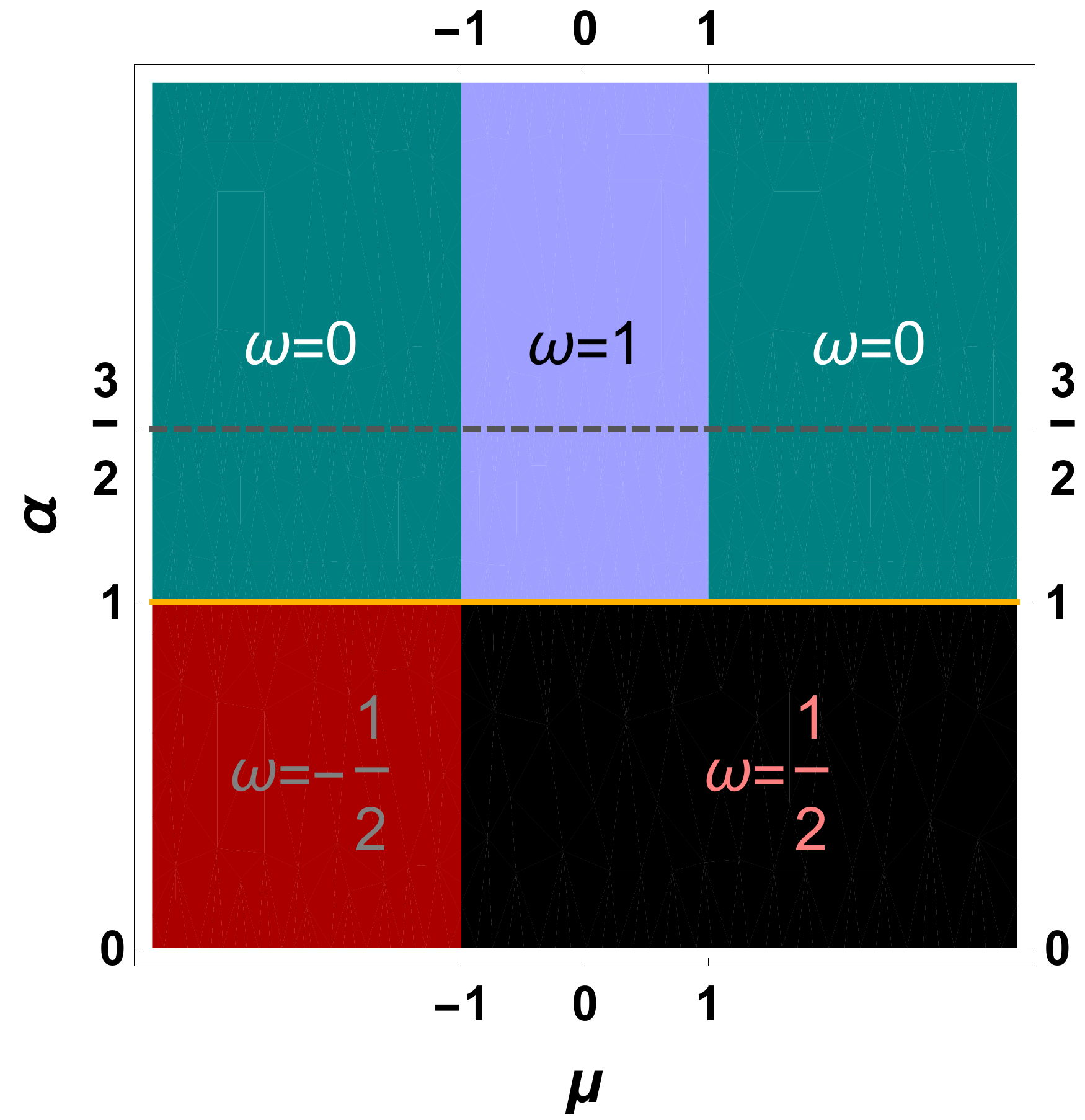}
\label{PD}}
\caption{(a) Energy spectrum $E$ of the LRK chain with open boundary 
conditions plotted versus the chemical potential $\mu$ for $\al = 0.7$ and 
system size $L = 1000$. We see that there are no end modes lying in the bulk 
energy gap at $\mu \leq -1$. (b) Phase diagram of the LRK chain in the $\mu - 
\al$ plane. For $\al>1$, the phase diagram is topologically equivalent to the 
SRK chain with three phases: two non-topological phases with $\nu = 0$ and a 
topological phase for $-1 < \mu < 1$ which has $\nu = 1$ and hosts MZMs at the 
ends of the system. For $\al<1$, the model has a non-topological 
phase for $\mu < -1$ with $\nu = -1/2$ and a topological phase for $\mu > -1$ 
which has $\nu = 1/2$ and hosts MDMs at the ends.} \label{UE2} \end{figure*}

\noi (b) We now focus our attention on the region $\al < 1$ which is truly an 
emergent feature of the long-range nature of the superconducting pairing; here
the phase diagram is expected to be drastically different from the conventional SRK model. In this region, for all $\mu <-1$ the system with open boundary 
conditions is in a trivial phase with no end modes, while for all $\mu > -1$, 
this system hosts topological MDMs at the ends (see Fig.~\ref{UEL}). 
These MDMs appear due to the coupling induced between the two MZMs at the 
two distant ends due to the presence of long-range pairing; hence 
the MDM formed is a highly non-local object. Moreover, although the Dirac 
mode is massive, it is still topological and is protected by the 
bulk gap. This non-local topological quasiparticle is also protected 
by the fermionic parity because the ground state of the system in this 
phase still retains its even parity; populating the MDM which is the first 
excited state of the system would then require a change in the fermionic parity 
from even to odd. Since no discrete symmetry has been broken by the 
inclusion of the long-range pairing, the system still belongs to the BDI 
symmetry class. However, the winding number $\nu$ is modified by the 
topological singularity at $k=0$ generated by the long-range pairing. This 
happens because at $k=0$ both the energy dispersion $E_k^{\pm}$ in 
Eq.~\eqref{eq_spectrum} and the group velocity $\pa_k E^{\pm}_k$ 
diverge as the integrand in the definition of the Berry/Zak phase in 
Eq.~\eqref{zak} is considered in the limit $k \to 0$. For the trivial phase 
with $\mu < -1$, the winding number is $\nu = -1/2$, whereas in the topological 
phase hosting MDMs which occurs for $\mu > -1$, it turns out that $\nu = +1/2$.
Although the topological invariant is a half-integer in both cases, the 
difference between the invariants in the two topologically different phases is 
unity, indicating that a topological phase transition separates the two 
half-integer quantized topological phases. The parameter range $\al <1$ is 
thus in perfect agreement with the results of Ref.~\onlinecite{viyuela16}.\\

\begin{figure*}[]
\centering
\subfigure[]{%
\includegraphics[width=.40\textwidth,height=4.3cm]{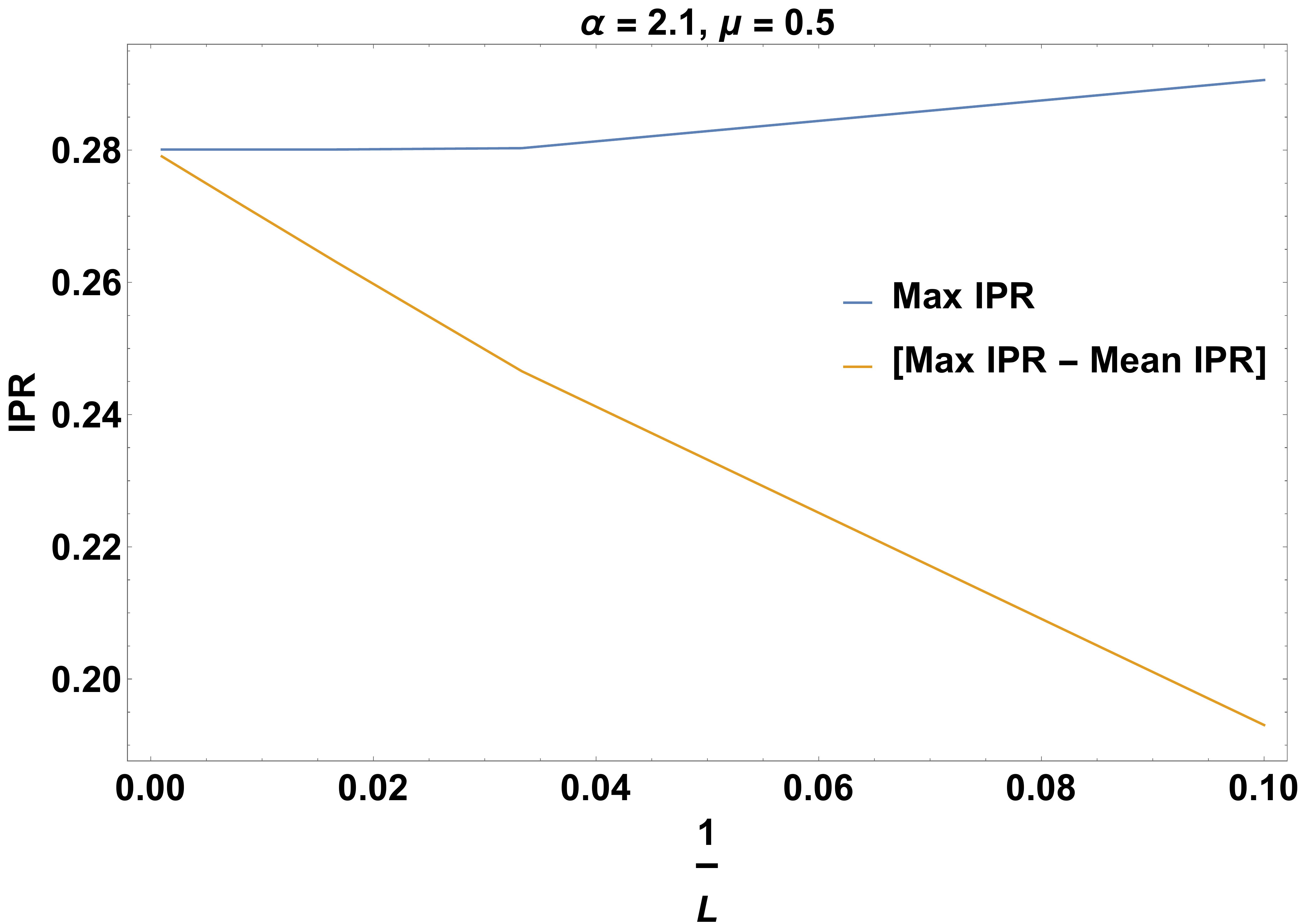}
\label{IPR1}}
\hfill
\quad
\subfigure[]{%
\includegraphics[width=.40\textwidth,height=4.3cm]{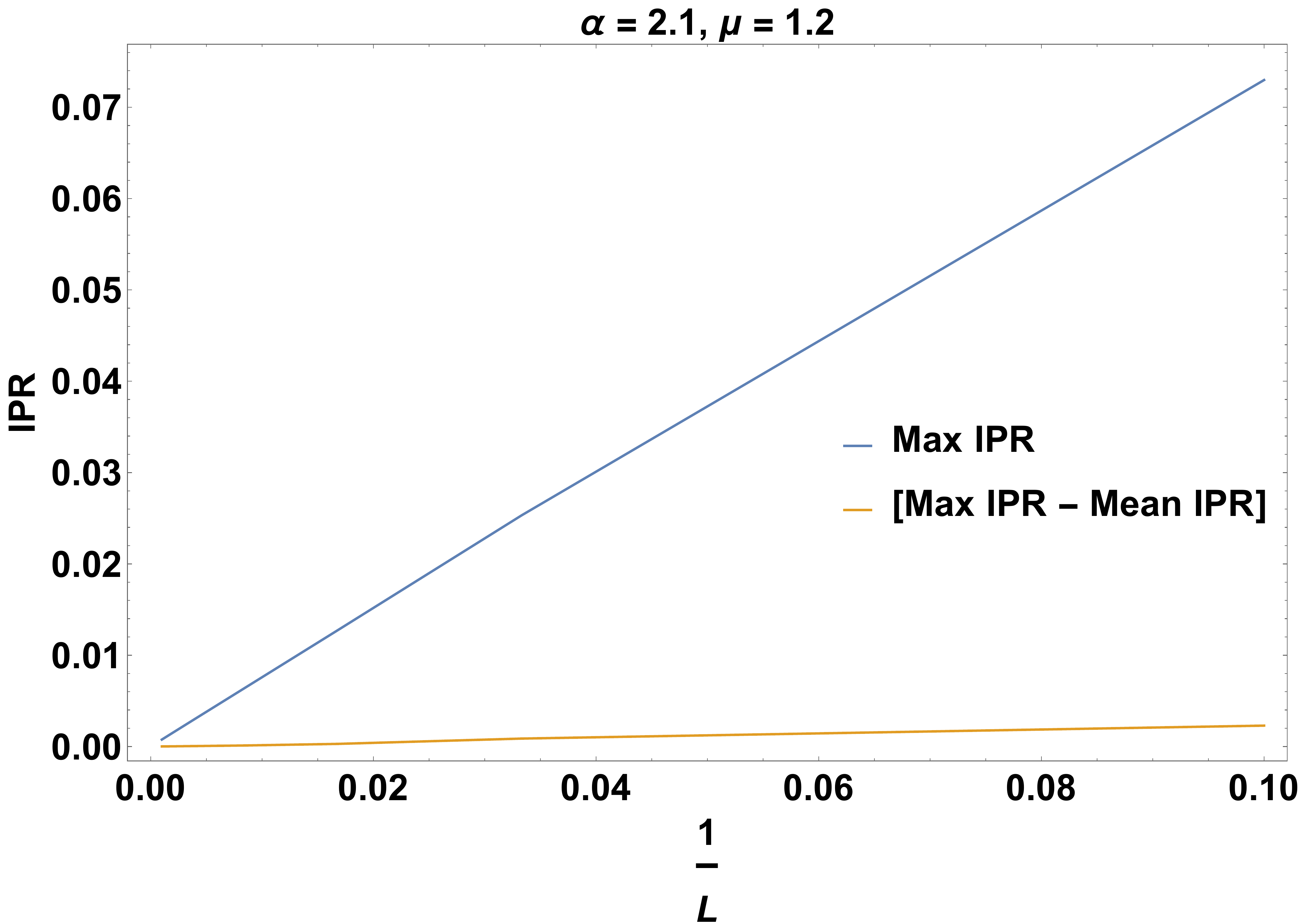}
\label{IPR2}}
\centering
\subfigure[]{%
\includegraphics[width=.40\textwidth,height=4.3cm]{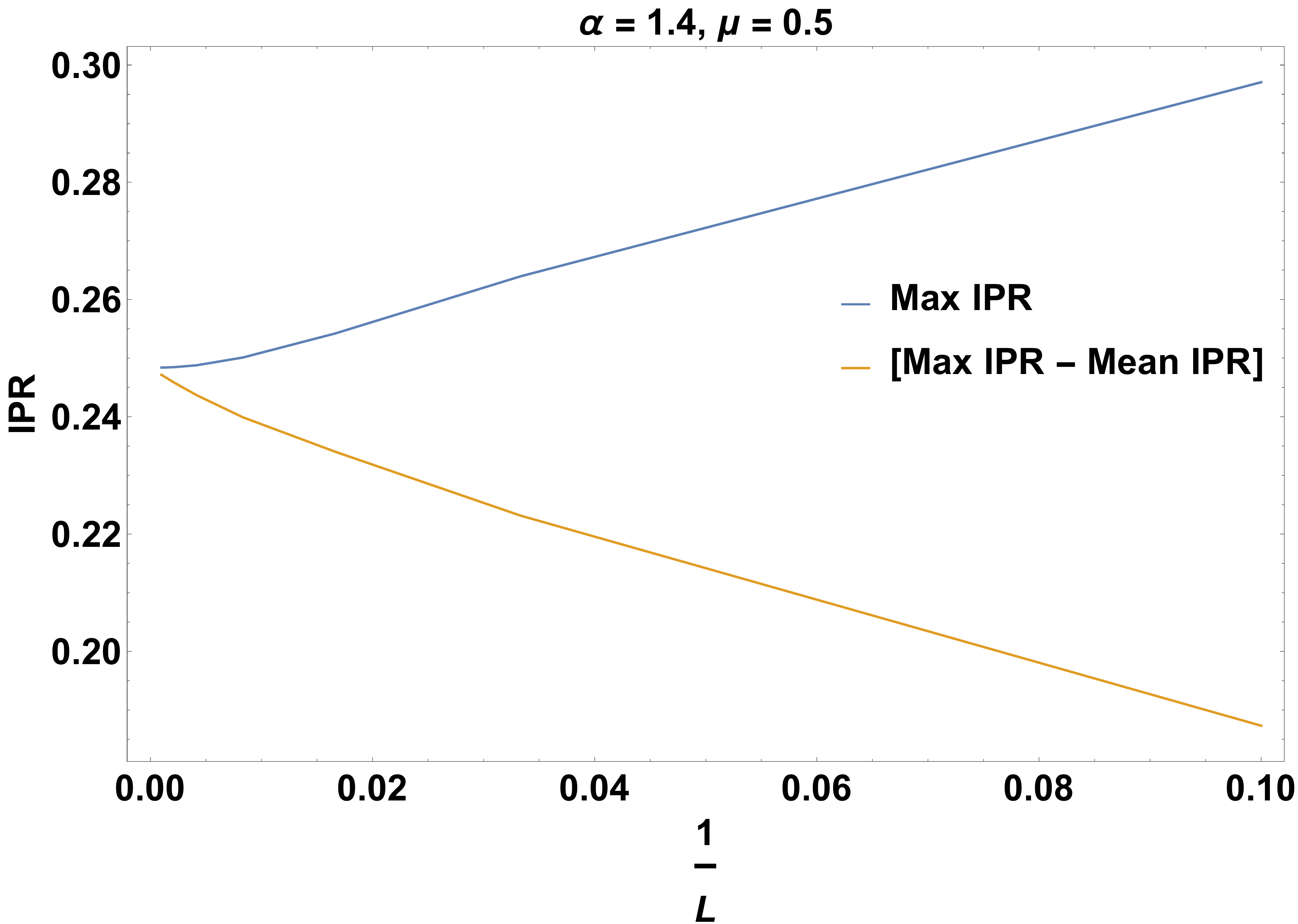}
\label{IPR3}}
\hfill
\quad
\subfigure[]{%
\includegraphics[width=.40\textwidth,height=4.3cm]{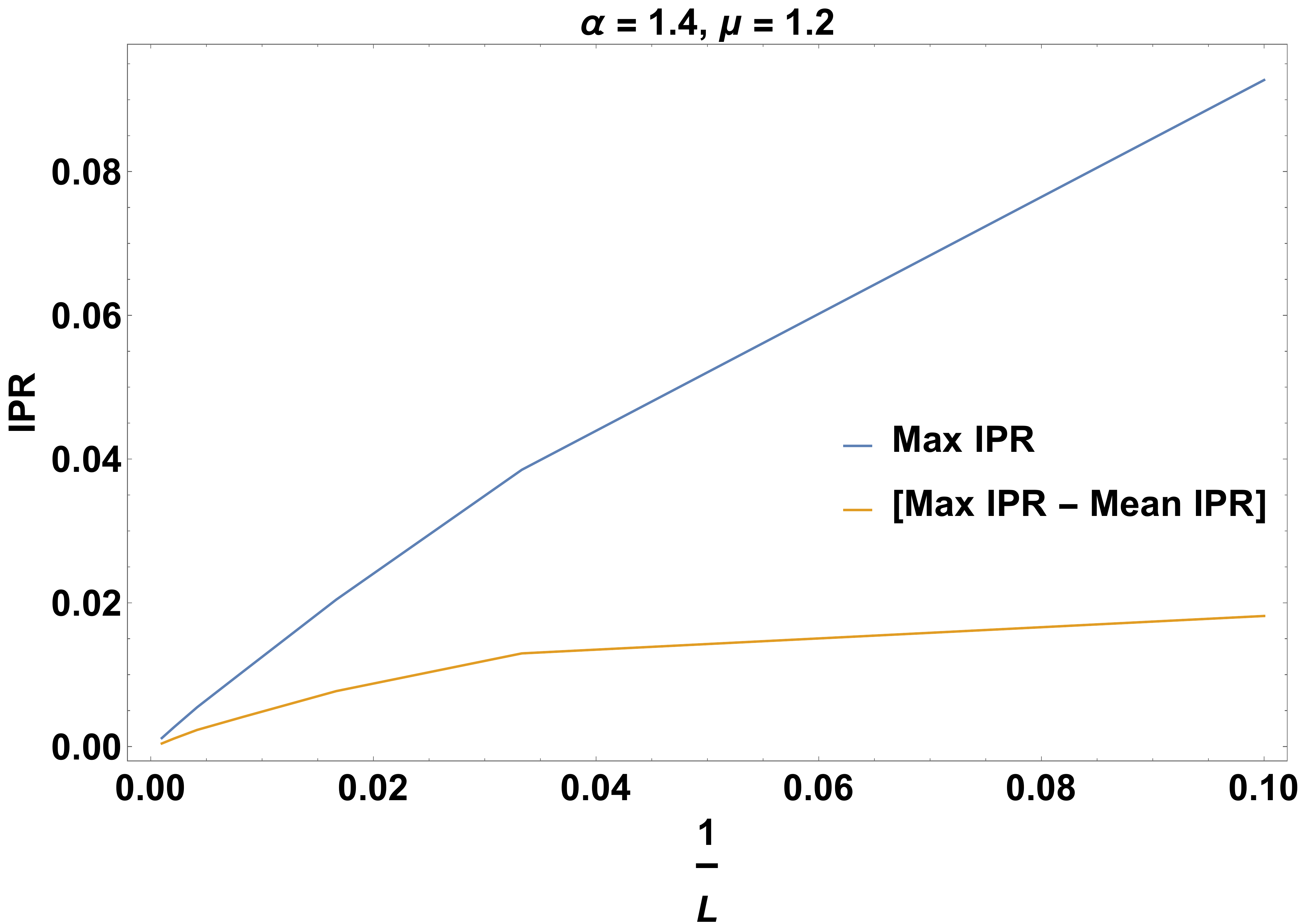}
\label{IPR4}}
\centering
\subfigure[]{%
\includegraphics[width=.40\textwidth,height=4.3cm]{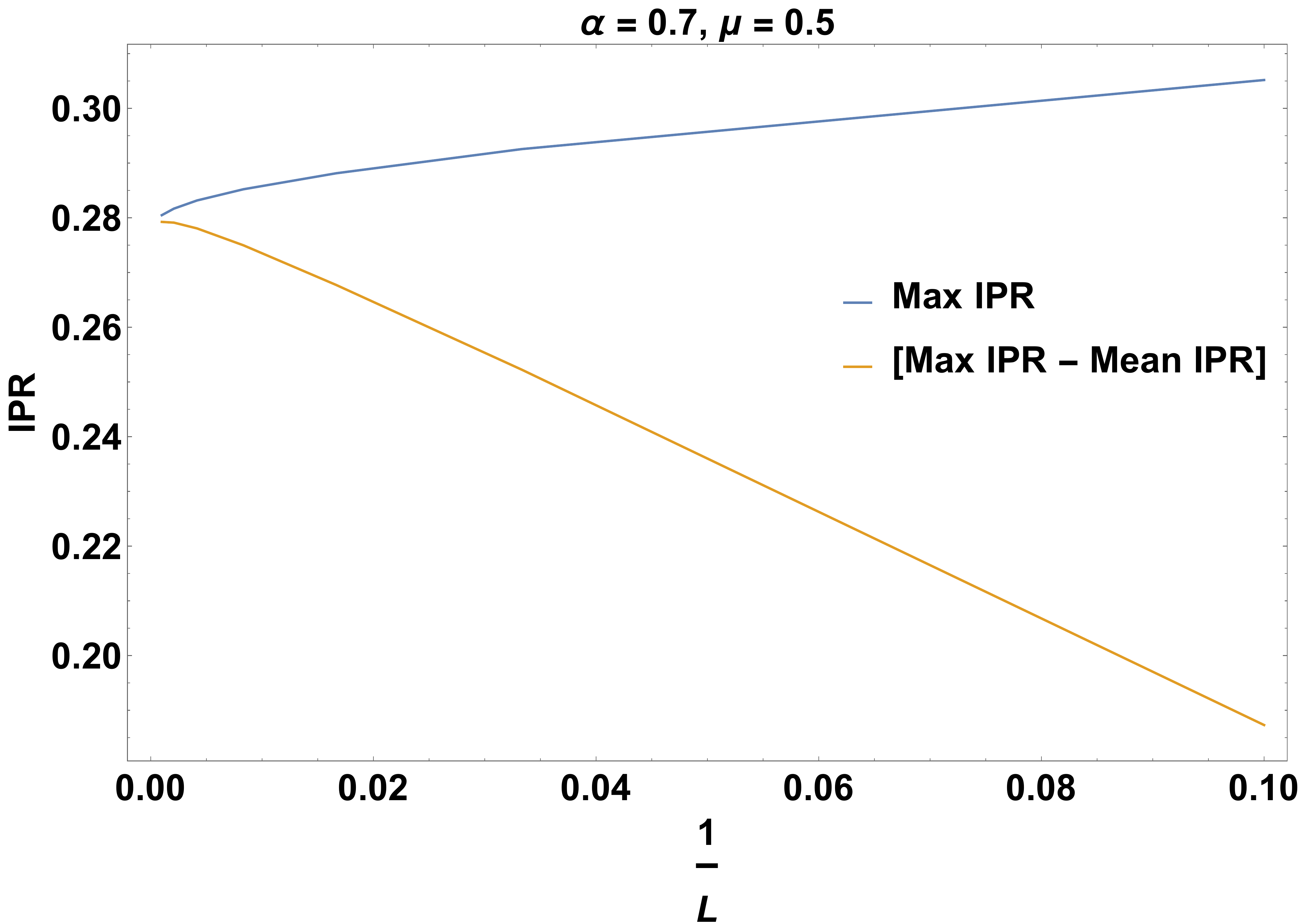}
\label{IPR5}}
\hfill
\quad
\subfigure[]{%
\includegraphics[width=.40\textwidth,height=4.3cm]{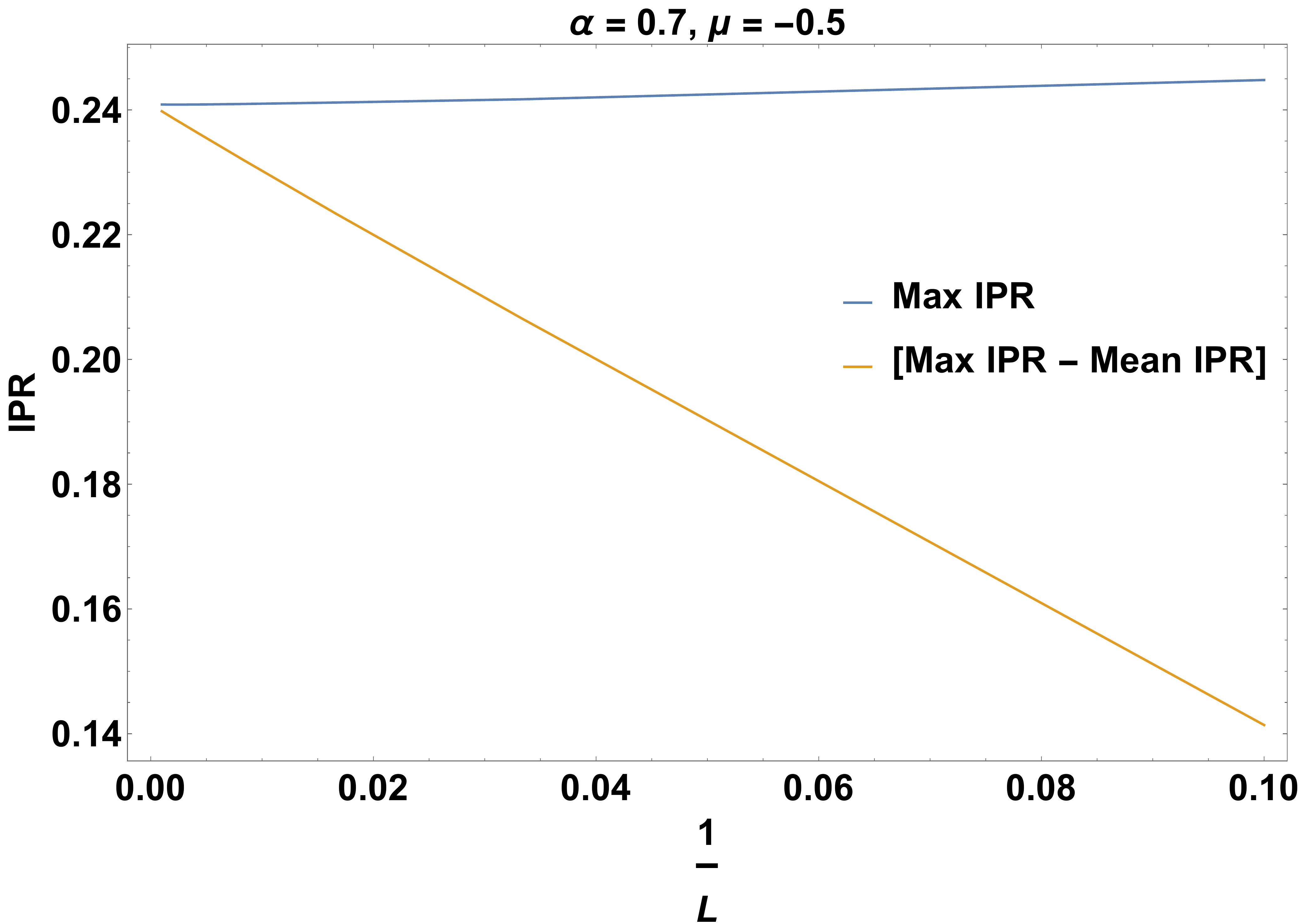}
\label{IPR6}}
\centering
\subfigure[]{%
\includegraphics[width=.40\textwidth,height=4.3cm]{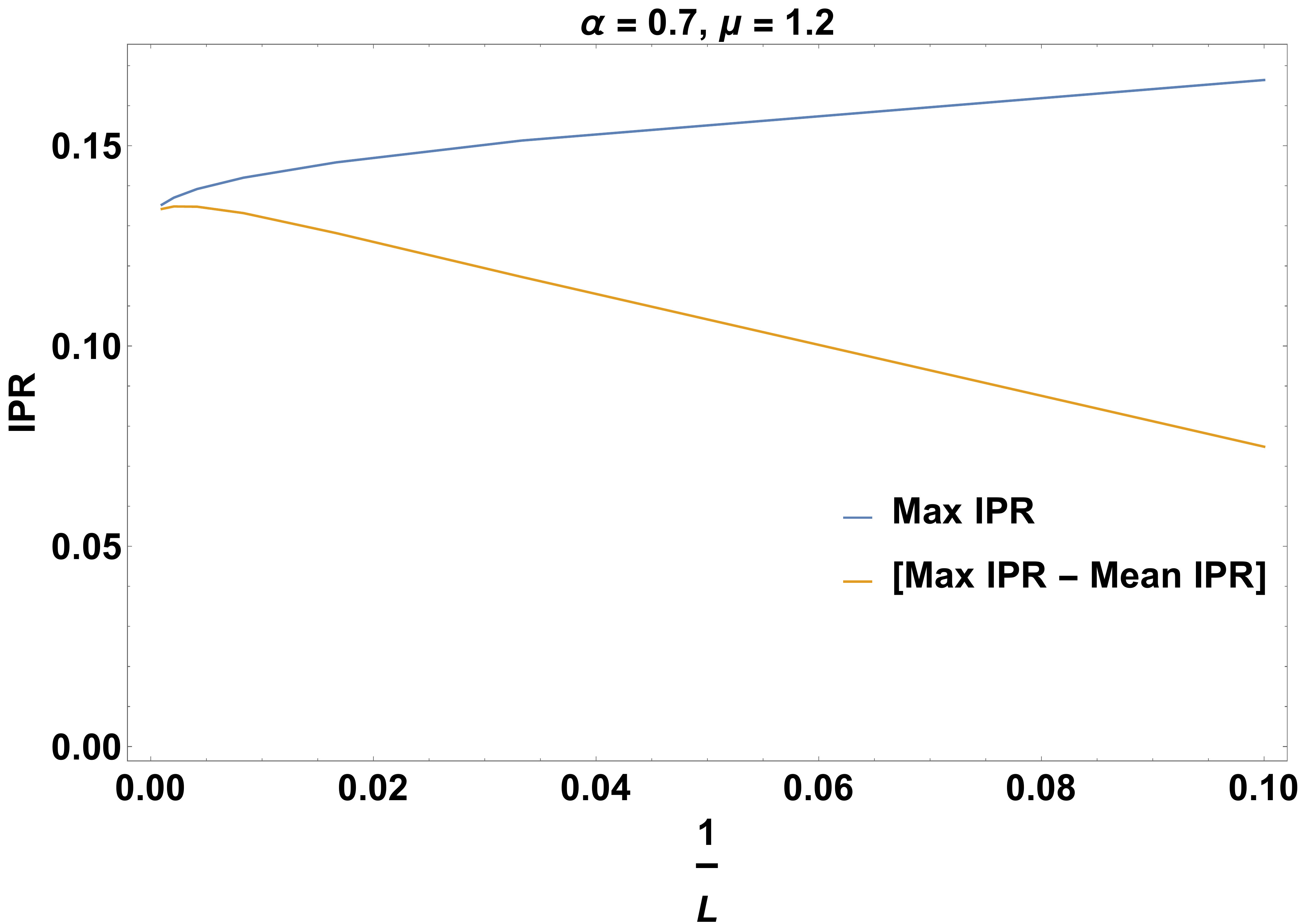}
\label{IPR7}}
\hfill
\quad
\subfigure[]{%
\includegraphics[width=.40\textwidth,height=4.3cm]{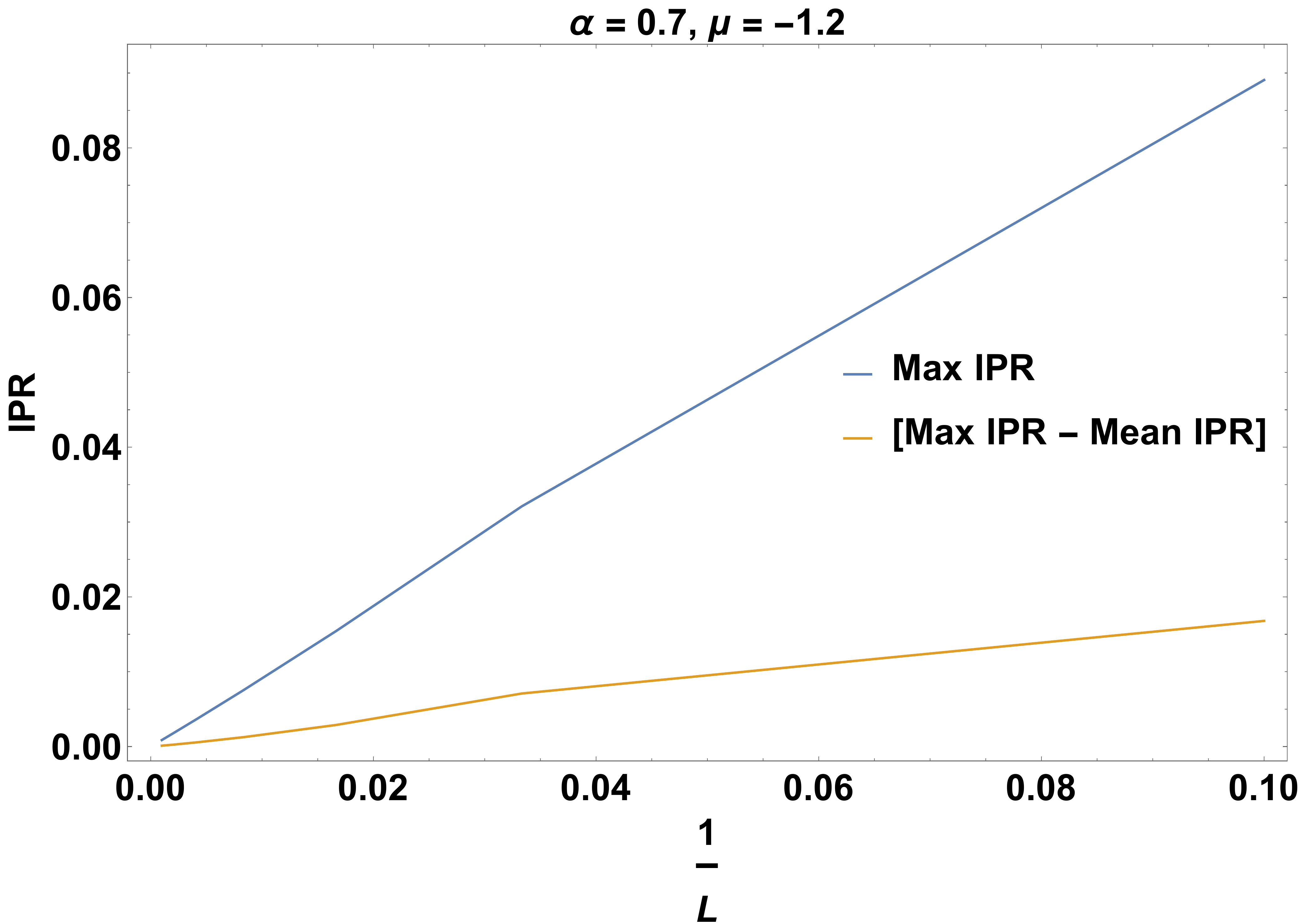}
\label{IPR8}}
\caption{Maximum value of the IPR (Max[$I_j$] in blue) and the difference 
between the maximum and mean values of the IPR ($\De I_j=$ Max[$I_j$]-
Mean[$I_j$] in yellow) plotted versus the inverse system size $1/L$. (a) 
For $\al = 2.1$ and $\mu = 0.5$, both Max[$I_j$] and $\De I_j$ converge to 
a finite non-zero value with decreasing $1/L$ indicating that localized end 
modes are present. (b) For $\al = 2.1$ and $\mu = 1.2$, both Max[$I_j$] 
and $\De I_j$ approach zero with decreasing $1/L$ indicating the absence of 
localized end modes. (c) For $\al = 1.4$ and $\mu = 0.5$ (these values
are selected from the apparent crossover region), both Max[$I_j$] and 
$\De I_j$ converge to a finite non-zero value with decreasing $1/L$ 
indicating that localized end modes are present. (d) For $\al = 1.4$ and 
$\mu = 1.2$, both Max[$I_j$] and $\De I_j$ approach zero with decreasing 
$1/L$ indicating the absence of localized end modes once again. 
For $\al = 0.7$, the presence of (massive) localized end modes have been 
confirmed for (e) $\mu = 0.5$, (f) $\mu = -0.5$ and (g) $\mu = 1.2$, by 
observing a finite non-zero value of Max[$I_j$] and $\De I_j$ as $1/L\to 0$, 
whereas localized end modes of any kind are absent for (h) $\mu = -1.2$ as 
indicated by their zero values as $1/L\to 0$.} \label{IPRG} \end{figure*}

We will now ascertain whether the energy eigenvectors whose energies lie in 
the gap of the bulk bands are localized or not. A convenient numerical method 
for checking this is to look at the inverse participation ratio (IPR) for 
different values of $\al$ and $\mu$. We assume that the eigenvectors of the 
open chain Hamiltonian in Eq.~\eqref{eq_hamm}, denoted as $\ket{e_j}$, are 
normalized so that $\sum_{m=1}^{2L} |e_j(m)|^2 = 1$ for each value of $j$; 
here $m= 1,2,... ,2L$ labels the Majorana components $a_m$ of the eigenvector 
$\ket{e_j}$. The IPR of an eigenvector is then defined as, $I_j = 
\sum_{m=1}^{2L} |e_j(m)|^4$.
If $\ket{e_j}$ is extended equally over all sites so that $|e_j (m)|^2 = 
1/(2L)$ for each $m$, then $I_j = 1/(2L)$ will approach zero in the 
thermodynamic limit. But if $\ket{e_j}$ is localized over a decay length 
$\xi$ and remains constant as $L \to \infty$, then $|e_j (m)|^2 \sim 1/\xi$ 
in a region of length $\xi$ and $\sim 0$ elsewhere. This yields $I_j \sim 
1/\xi$ which will remain finite as $L \to \infty$. If $L$ is 
sufficiently large, a plot of $I_j$ versus $j$ will be able to distinguish 
between states which are localized (over a length scale $\ll L$) and states 
which are extended. Therefore, to identify whether localized states exist in 
the various topological phases of the LRK chain, we plot 
the maximum value of the IPR, Max[$I_j$], amongst all the states
$j$ versus the inverse of the system size ($1/L$); this is shown in
Fig.~\ref{IPRG}. In the same figure, we also show how the difference 
between the maximum and the mean of the IPR, Max[$I_j$] ~-~ Mean[$I_j$], 
scales with $1/L$. If both Max[$I_j$] and Max[$I_j$] ~-~ Mean[$I_j$] 
converge to finite non-zero values which are close to each other with 
increasing system size $L$, it confirms the presence of a localized mode. 
In all such cases, we have verified numerically that the maximum value 
of the IPR coincides with the IPR of an eigenstate whose energy eigenvalue 
lies within the bulk gap of the system. On the other hand, if both Max[$I_j$] 
and Max[$I_j$] ~-~ Mean[$I_j$] approach zero as $1/L \to 0$, it confirms the 
absence of any localized modes in the system. \\

To summarize, the phase diagram (see Fig.~\ref{PD}) of the LRK chain 
consists of two regions depending on the value of $\al$: (i) the region
$\al>1$ is effectively short-ranged and hosts massless MZMs for $-1<\mu<1$ 
where the winding number is $\nu = 1$, and (ii) the region $\al<1$ where 
the effects of long-range pairings couple the two end modes together for 
all $\mu > -1$, thereby generating topologically robust MDMs. Interestingly, 
this phase is characterized by a winding number $\nu = 1/2$. The system lies 
in a topologically trivial phase with no end modes for $\al > 1$ and $|\mu|>1$ 
where the winding number is zero, and also for $\al < 1$ and $\mu < -1$ 
where the winding number is $-1/2$.

\section{Floquet evolution: periodic $\de$-function kicks in chemical 
potential}
\label{sec4}

In this section, we consider what happens when the chemical potential $\mu$ 
is given $\de$-function kicks periodically in time. One reason for choosing 
to consider periodic kicks is that this is known to produce interesting 
effects in quantum systems such as dynamical localization~\cite{stock}. We 
will also see that the effects of periodic kicks are considerably easier to 
study both numerically and analytically compared to the case where $\mu$ 
varies harmonically with time. We will take the chemical potential in 
Eq.~\eqref{eq_ham} to be of the form
\beq \mu (t) ~=~ \mu ~+~ V \sum_{n=-\infty}^\infty \de (t - nT), 
\label{ht1} \eeq
where $V$ is the kicking strength, $T=2\pi/\om$ is the time period and $\om$ 
is the frequency of kicking. Eqs.~\eqref{eq_ham} and \eqref{ht1} imply that 
the system has time-reversal symmetry: $H^* (-t) = H(t)$ for all values
of $t$. (In general, we note that a system is said to have time-reversal 
symmetry if we can find a time $t_0$ such that $H^* (t_0 - t) = H(t)$ for 
all $t$, and does not have time-reversal symmetry if no such $t_0$ exists). 
As discussed below, we numerically compute the time evolution operator 
$U(T,0)$ for periodic kicking for various values of the parameters $\ga$, 
$\De$, $\al$, $\mu$, $V$, $\om$ and the system
size $L$. We then find all the eigenvalues and eigenvectors of $U(T,0)$.
Since the system is invariant under parity $\cal P$ (corresponding
to reflecting the system about its mid-pont), we can choose the
eigenvectors of $U(T,0)$ to also be eigenvectors of $\cal P$.

The bulk Floquet operator $U_F$ for periodic $\de$-function kicks can be 
written as a product of three terms: a kicked evolution with a chemical 
potential $V \de (t-T)$ followed by an evolution via the free Hamiltonian 
$H_k$ with a constant chemical potential $\mu$ for time $T$ and then again 
a kicked evolution with a chemical potential $V \de (t-T)$. This symmetric 
choice of kicking makes the particle-hole, the chiral and the time-reversal 
symmetries more transparent and yields the Floquet operator $U_F$ for each 
momentum mode $k$ as~\cite{thakurathi}
\beq U^F_k ~=~ U_k(T,0) ~=~ e^{-iV\si^z} ~e^{-i H_k T} ~e^{-iV\si^z}. 
\label{eq_flo2x} \eeq
The Floquet Hamiltonian $H^F_k$ can now be obtained from $U^F_k$ as
\beq \label{eq_hamf} H^F_k ~=~ \frac{i}{T}\ln{U^F_k} ~\equiv~ E^F_k 
\hat{c}_k \cdot \vec{\si}, \eeq
where the Floquet quasienergy $E^F_k$ defined within the Floquet Brillouin 
zone $-\pi/T$ to $\pi/T$ and the unit vector $\hat{c}_k$ are given by
\begin{widetext}
\bea \label{eq_fspec}
E^F_k &=& \frac{1}{T} ~\arccos \left[\cos\left(2V\right)\cos\left(E_k T
\right) ~-~ \frac{a_3}{E_k}\sin\left( 2V\right) \sin\left(E_k T\right)
\right], \non \\ \\ 
\hat{c}_k &=& \frac{1}{\sin\left(E^F_k T\right)} ~\Bigg\{\frac{a_2}{E_k}
\sin\left(E_k T\right)\hat{y} ~+ \left[\sin\left(2V\right)\cos\left(E_k T
\right) -\frac{a_3}{E_k}\cos\left( 2V\right) \sin\left(E_k T\right)\right]
\hat{z} \Bigg\}, \eea
\end{widetext}
where $a_2 = -f_\al (k)$ and $a_3 = \cos(k)-\mu$.

The same symmetrized Floquet evolution can also be carried out for the 
real-space Hamiltonian, but here the Floquet eigenvalues and eigenvectors 
have to be obtained numerically from the real-space Floquet operator 
$U^F_r = U_r(T,0)$. We also note that the eigenvalues of the unitary 
operator $U_r (T,0)$ for both open and periodic boundary conditions are given 
by phases, $e^{i\ta_j}$, and these must come in complex conjugate pairs if 
$e^{i\ta_j} \ne \pm 1$~\cite{thakurathi}. 
This is because $U_r (T,0)$ is a real operator when written in terms of
the Majorana operators $a_m$ in Eq.~\eqref{eq_hamm}. Hence $U_r (T,0) \psi_j = 
e^{i\ta_j} \psi_j$ implies that $U_r (T,0) \psi_j^* = e^{-i\ta_j} \psi_j^*$.

\section{Results}
\label{sec5}

We can see from Eq.~\eqref{eq_fspec} that the bulk gap for the Floquet 
spectrum can close (i.e., $E^F_k T$ can be equal to 0 or $\pi$) only at 
$k=0$ or $\pi$. Moreover, $E^F_k T$ in Eq.~\eqref{eq_fspec} can close at 
$E^F_k T = 0$ and $\pi$ for $k = 0$ only when $\al>1$ and for $k = \pi$ for
any value of $\al$. Therefore, to investigate the possible topological phase 
boundaries, we solve $E^F_k T = 0, \pi$ at $k=0, \pi$, for various values 
of $\mu$ keeping the values of the kicking frequency $\om = 2\pi/T$ and 
strength $V$ fixed. In the rest of this section, we will refer to $E^F_k T$, 
rather than $E^F_k$, as the Floquet quasienergy; we note that $E^F_k T$ is 
dimensionless and lies in the range $[-\pi,\pi]$.

When $\al>1$, we find that $E^F_k T = 0$ can occur if
\beq \label{eq_me}
\mu ~=~ \pm 1-\frac{\om}{\pi}\left[V-n\pi\right], \eeq
where $n = 0,\pm 1, \pm 2, \cdots$, and the $\pm$ in Eq.~\eqref{eq_me} stands 
for modes with $k = 0$ and $\pi$ respectively. On the other hand, for 
$\al<1$, although $E^F_k T = 0$ for $k = \pi$ for the same values of $\mu$ 
as above, $E^F_k T$ never crosses $0$ at $k=0$ due to the divergence of 
the polylogarithm function $f_\al (k)$ at that value of $k$.

Similarly, when the gap in the Floquet quasienergy $E^F_k T$ closes at 
$E^F_k T= \pi$, we obtain equations for the critical values of $\mu$ exactly 
as above. When $\al>1$, we find that $E^F_k T = \pi$ for
\beq \label{eq_mo}
\mu ~=~ \pm 1-\frac{\om}{\pi}\left[V-\left(n+\frac{1}{2}\right)\pi\right];~~~
n = 0,\pm 1, \pm 2, \cdots, \eeq
where $n = 0,\pm 1, \pm 2, \cdots$, and the $\pm$ in Eq.~\eqref{eq_mo} stands 
for modes with $k = 0$ and $k = \pi$ respectively. For $\al<1$, although $E^F_k
T = \pi$ for $k = \pi$ for the same values of $\mu$ as above, $E^F_k T$ never
becomes $\pi$ at $k = 0$ due to the divergence of the polylogarithm 
function $f_\al (k)$ at that value of $k$.

\begin{figure*}[]
\centering
\subfigure[]{%
\includegraphics[width=0.75\textwidth,height=4.9cm]{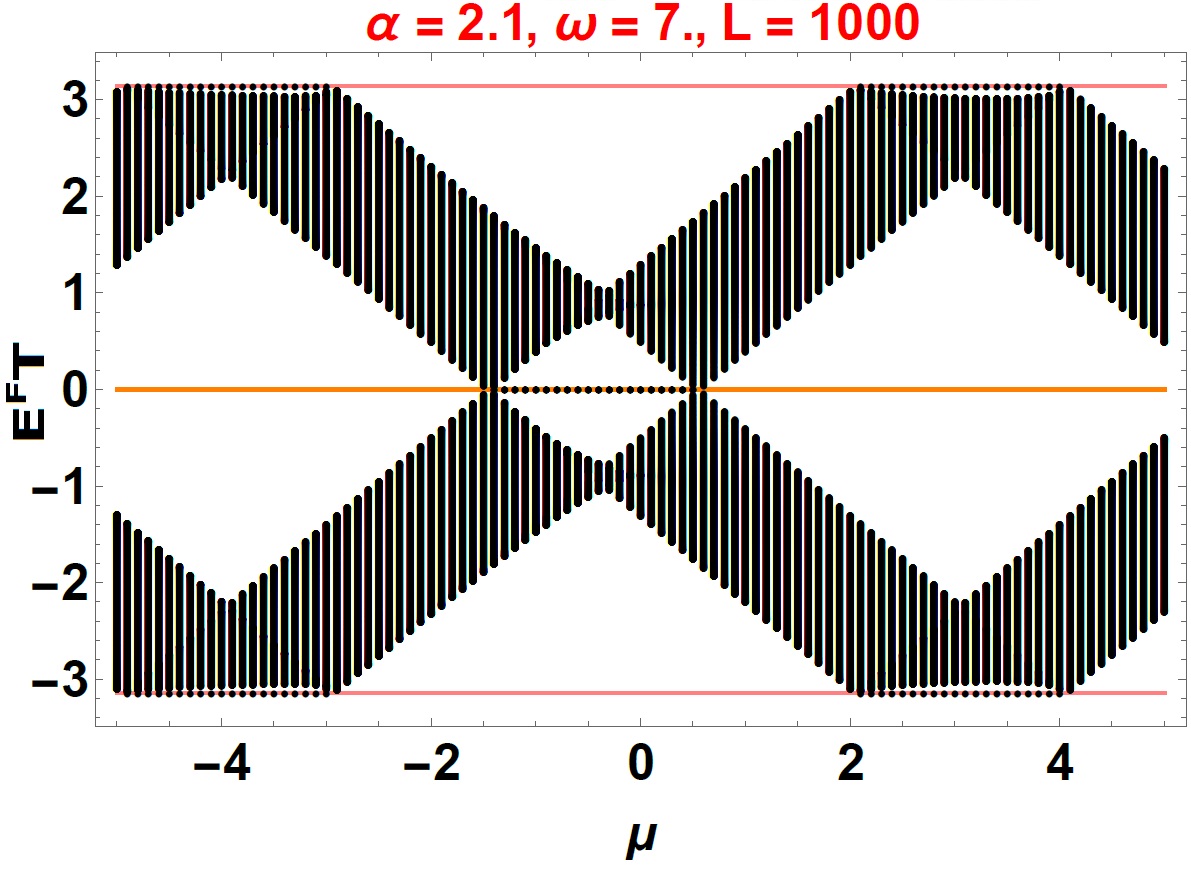}
\label{fsws1}}
\hfill
\quad
\subfigure[]{%
\includegraphics[width=0.75\textwidth,height=4.9cm]{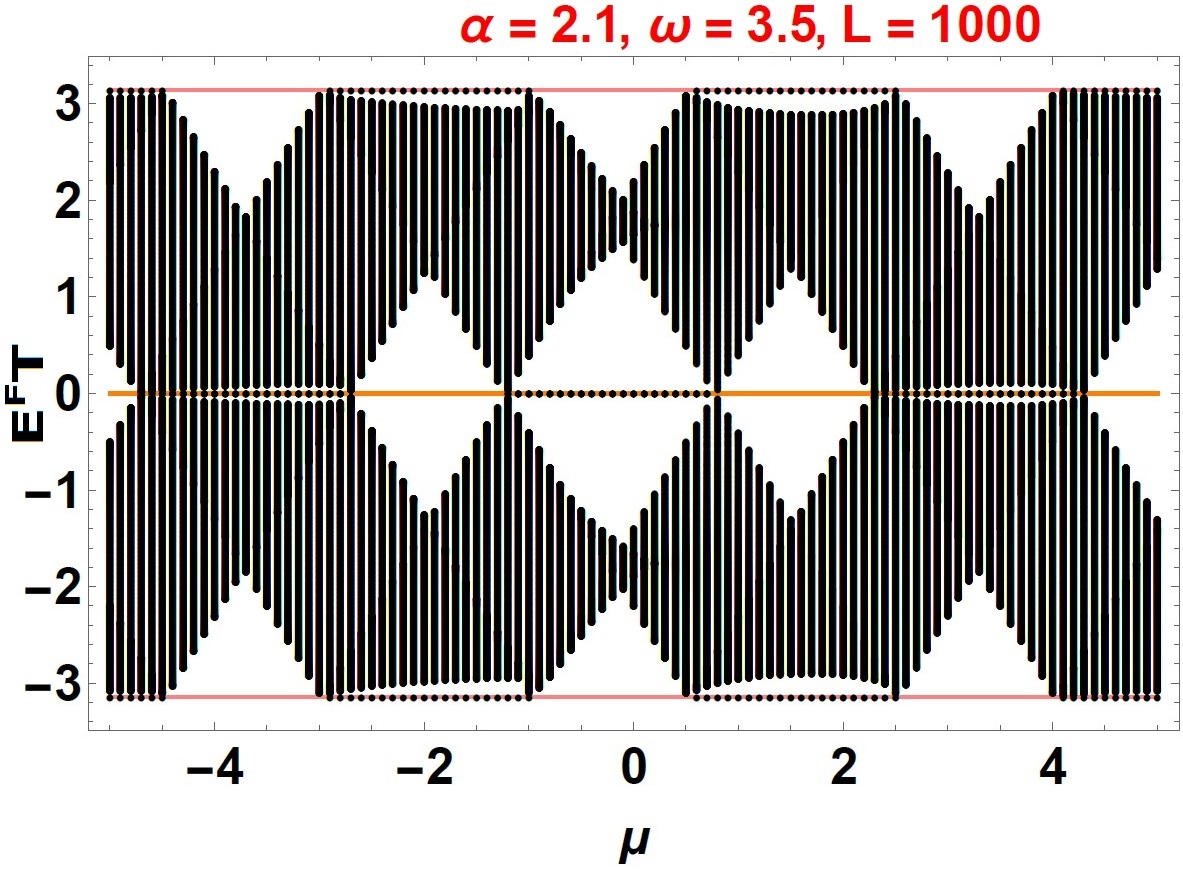}
\label{fsws2}}
\centering
\subfigure[]{%
\includegraphics[width=0.75\textwidth,height=4.2cm]{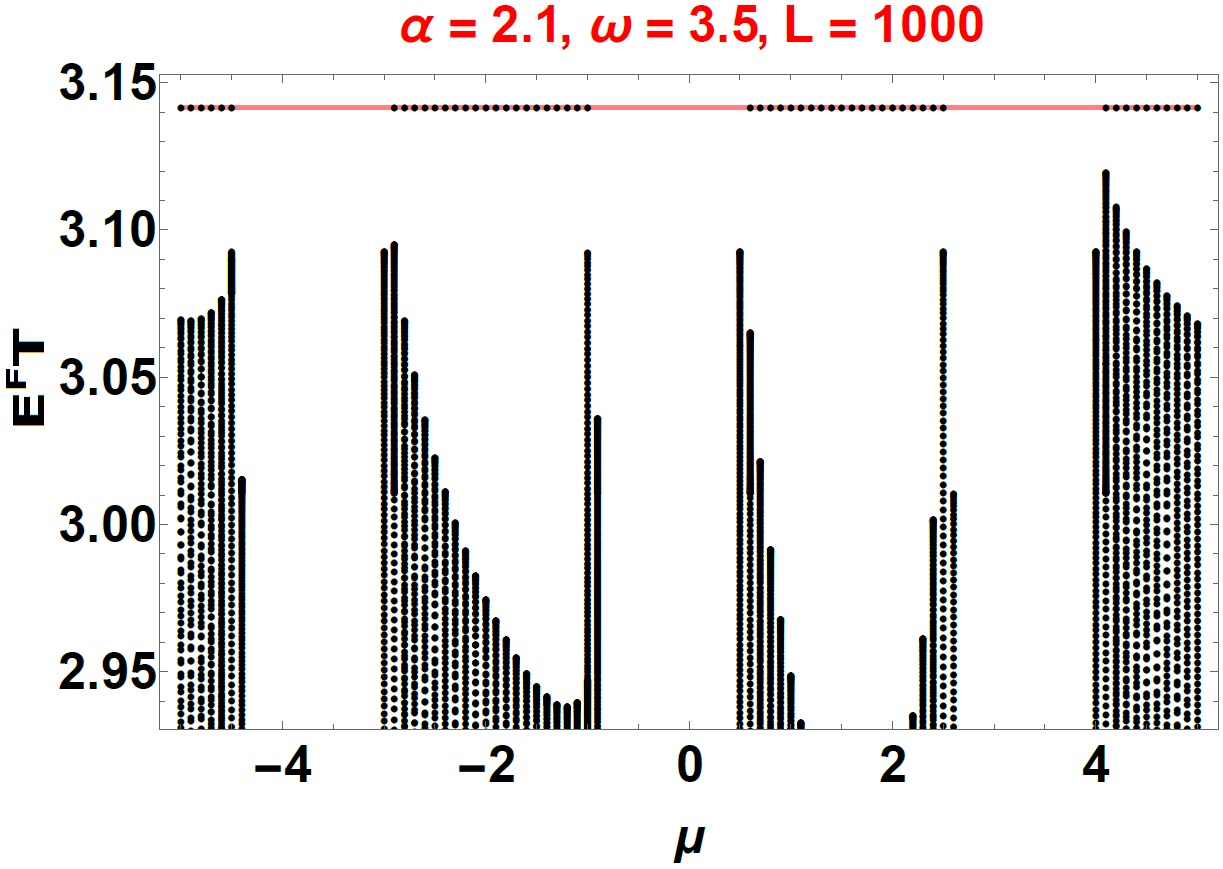}
\label{fsws2zt}}
\hfill
\quad
\subfigure[]{%
\includegraphics[width=0.75\textwidth,height=4.2cm]{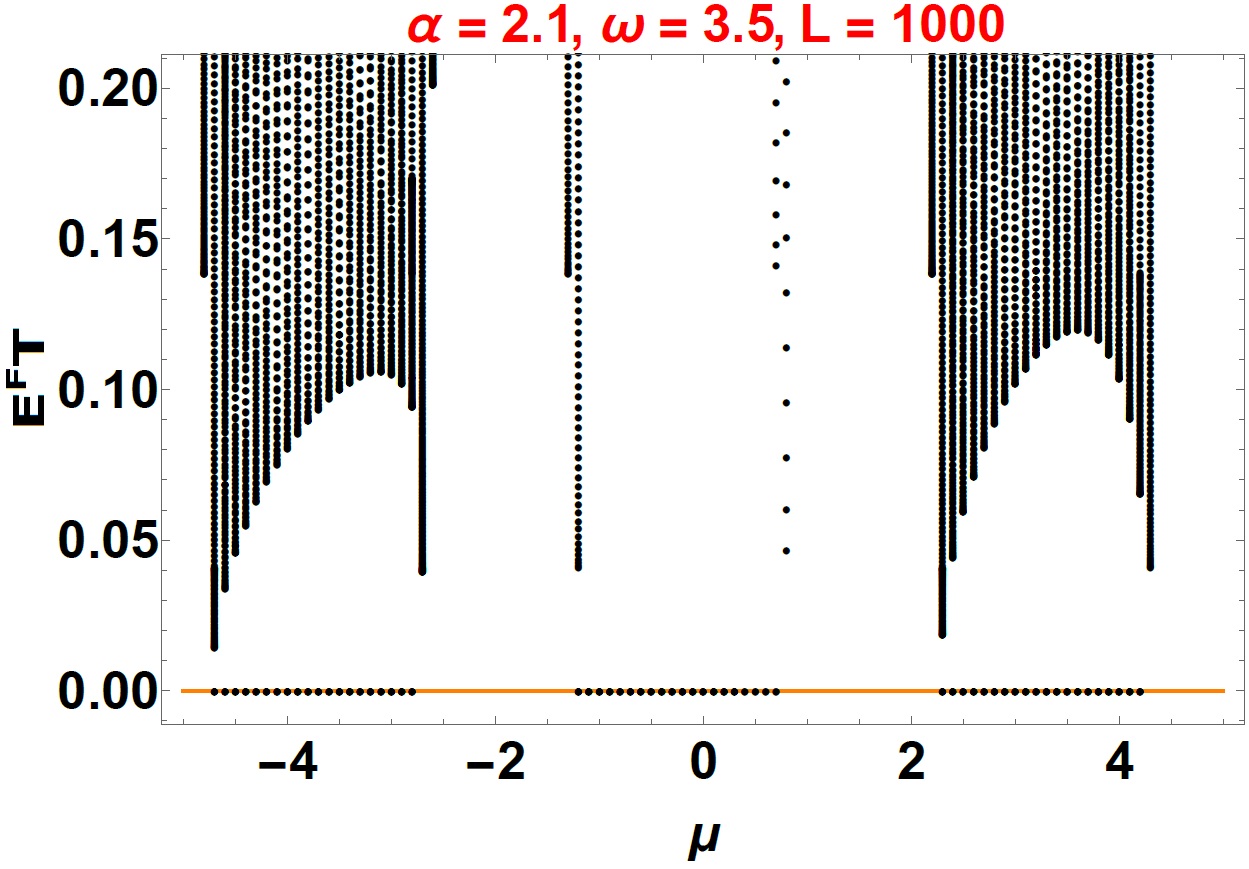}
\label{fsws2zb}}
\caption{Floquet quasienergy spectrum $E^F T$ of the LRK chain with open 
boundary conditions plotted versus the chemical potential $\mu$ for $\al = 
2.1$, kicking strength $V = 0.2$, kicking frequencies $\om = 7.0$ and $3.5$, 
and system size $L = 1000$. (a) For $\om = 7.0$, we see that
massless end modes are present at $E^F T = 0$ and $\pi$ for 
certain ranges of values of $\mu$ where there are no end modes in the undriven 
energy spectrum. However, end modes at $E^F T = 0$ and $\pi$ do not appear 
together at any value of $\mu$. (b) For $\om = 3.5$, massless end modes are 
present at $E^F T = 0$ and $\pi$ for certain ranges of $\mu$ which are 
different from (a). We see here that massless end modes at $E^F T = 0$ and 
$\pi$ can appear together for some values of $\mu$, for instance, $\mu = 
- 0.9$. For $\om = 3.5$, the quasienergy spectrum has been zoomed around 
(c) $E^F T=\pi$ and (d) $E^F T = 0$. They show that the modes are truly 
massless, and also that massless end modes at $E^F T = 0$ and $\pi$ can 
co-exist at certain values of $\mu$, such as $\mu = -0.9$.} \label{fig_fss} 
\end{figure*}

\begin{figure*}[]
\centering
\subfigure[]{%
\includegraphics[width=0.75\textwidth,height=4.9cm]{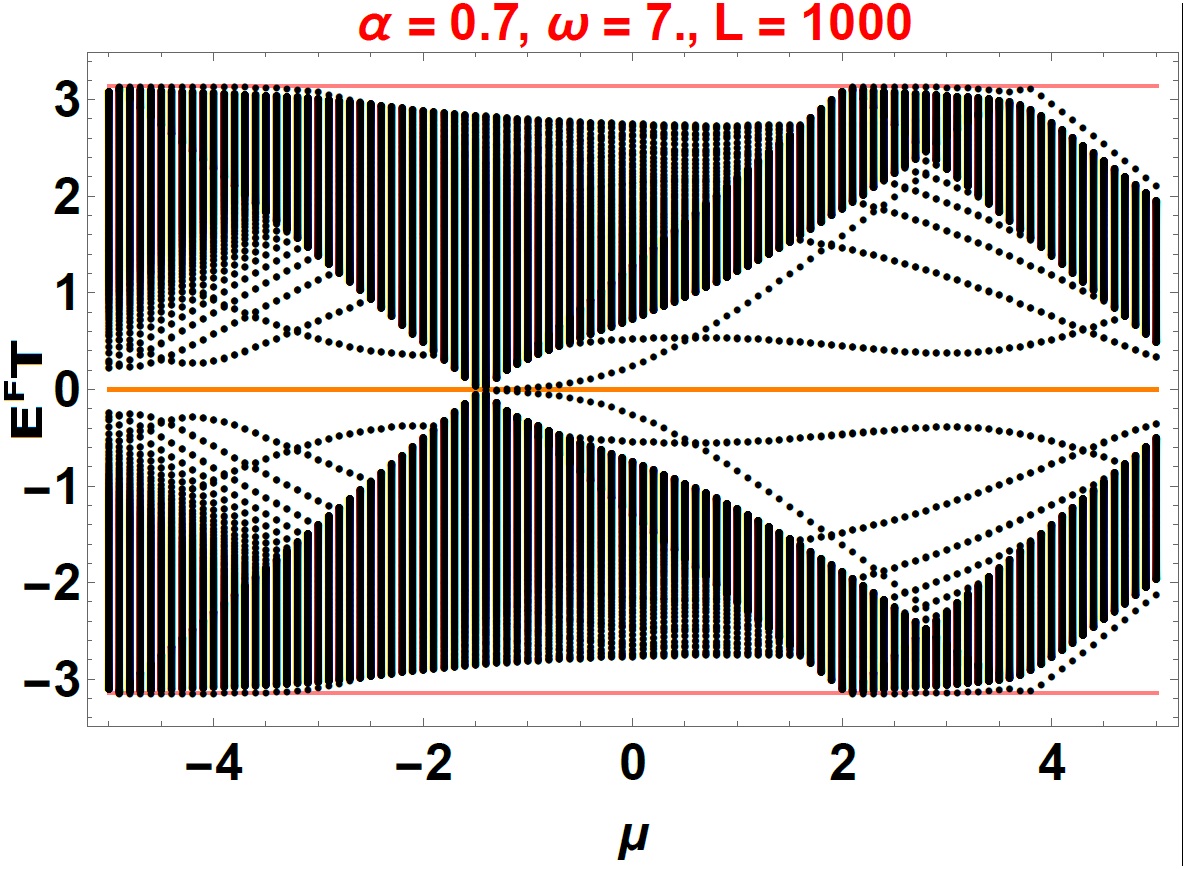}
\label{flsw1}}
\hfill
\quad
\subfigure[]{%
\includegraphics[width=0.75\textwidth,height=4.9cm]{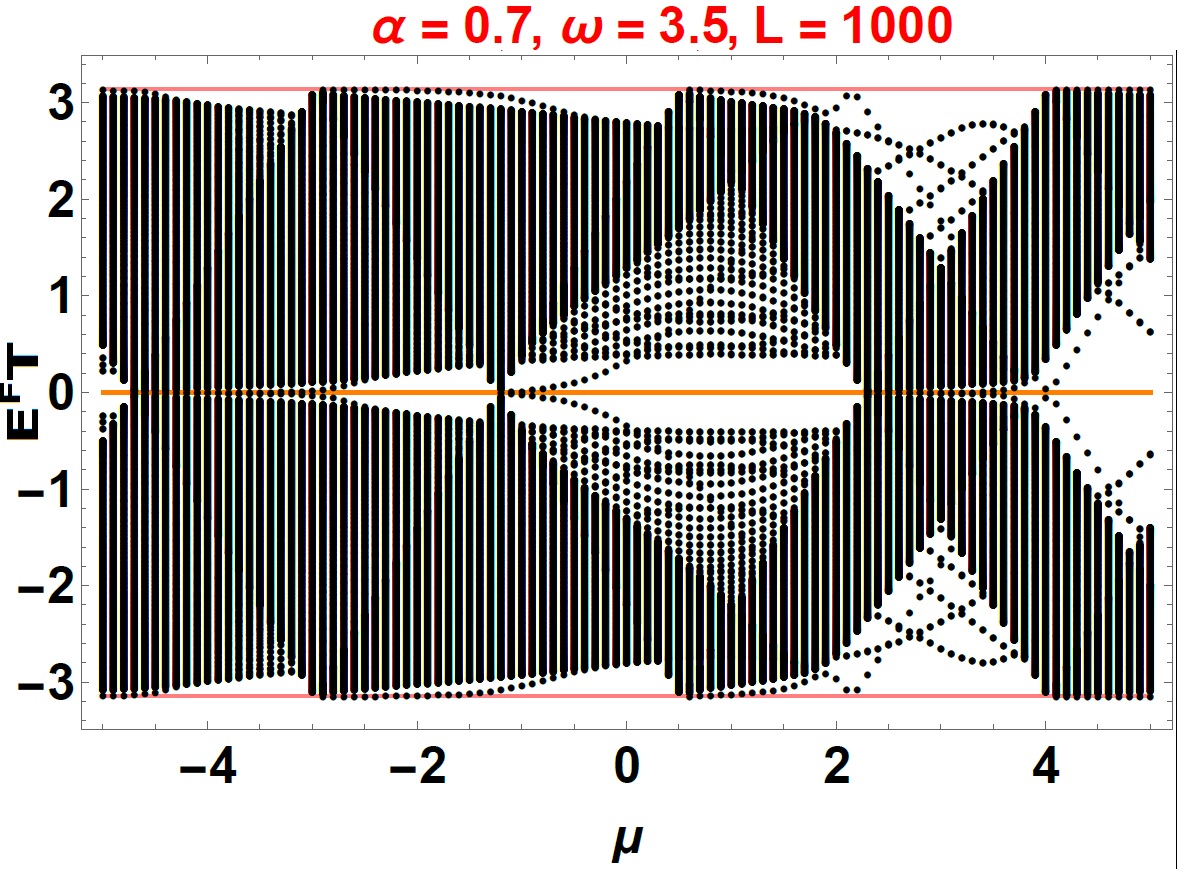}
\label{flsw2}}
\centering
\subfigure[]{%
\includegraphics[width=0.75\textwidth,height=4.2cm]{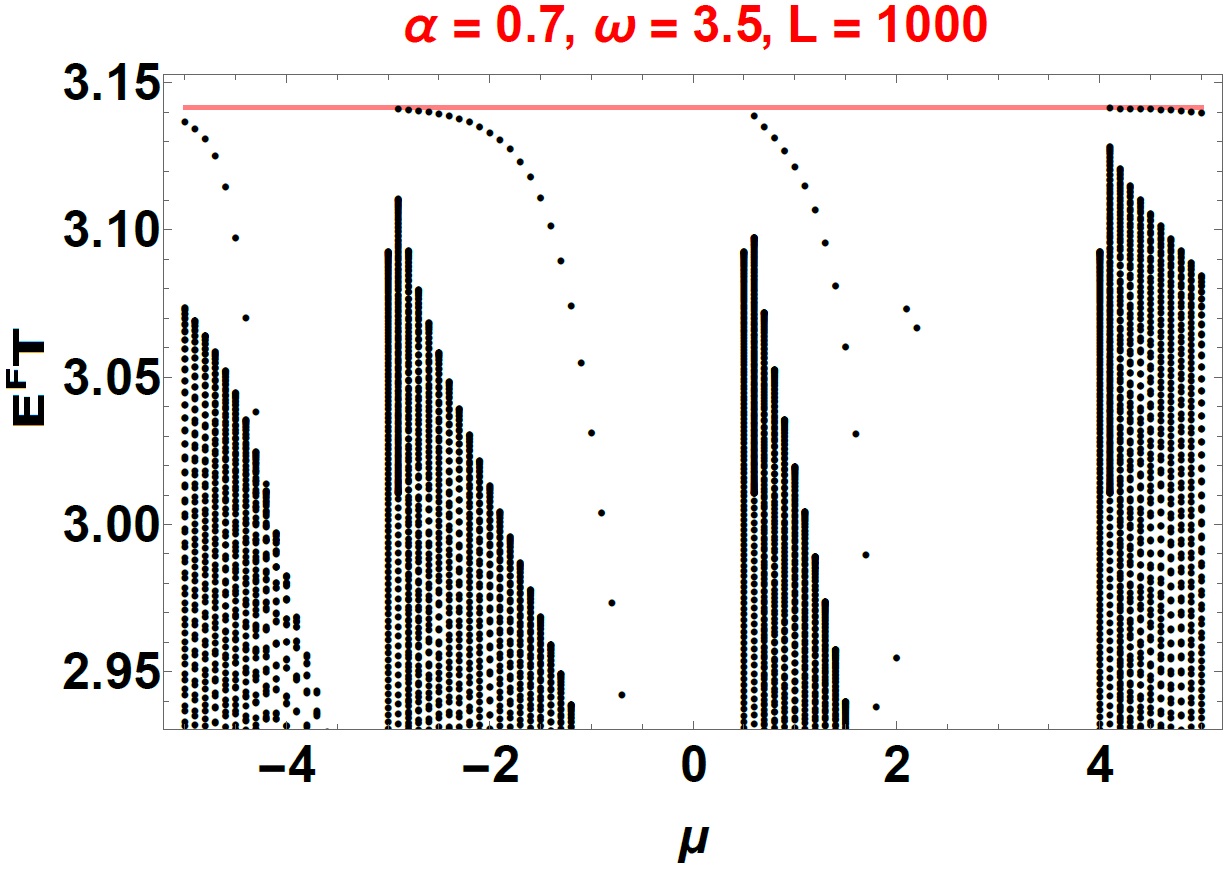}
\label{flw2zt}}
\hfill
\quad
\subfigure[]{%
\includegraphics[width=0.75\textwidth,height=4.2cm]{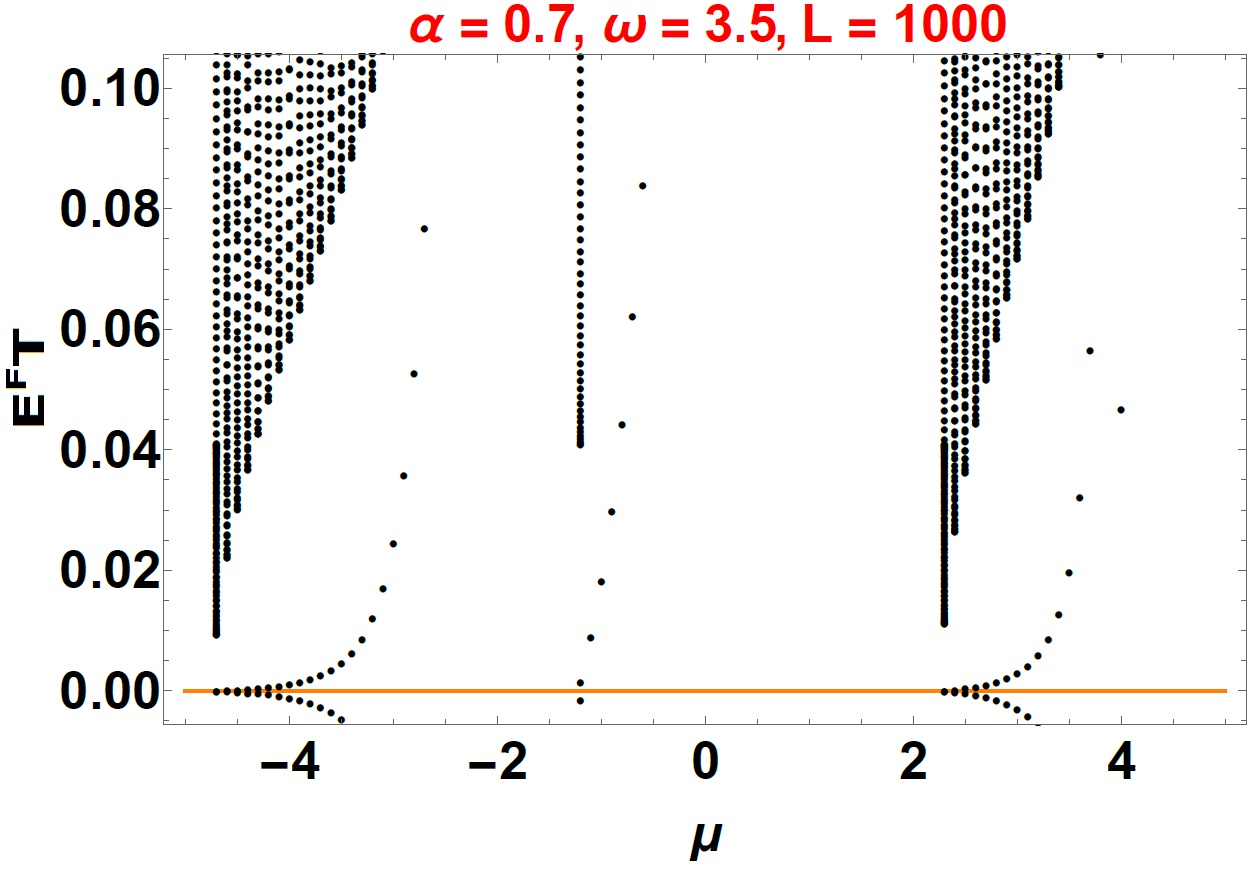}
\label{flw2zb}}
\caption{Floquet quasienergy spectrum $E^F T$ of the LRK chain with open 
boundary conditions plotted versus $\mu$ for $\al = 0.7$, kicking strength 
$V = 0.2$, kicking frequencies $\om = 7.0$ and $3.5$, and system size $L = 
1000$. (a) For $\om = 7.0$, we see that massive end modes are present around 
$E^F T = 0$ and $\pi$ for certain ranges of values of $\mu$ where there are no 
end modes in the undriven energy spectrum. (b) For $\om = 3.5$, massive end 
modes are present around $E^F T = 0$ and $\pi$ 
for certain ranges of $\mu$ which are different from (a). The occurrence 
of massive end modes around $E^F T = 0$ and $\pi$ can now appear together for 
some values of $\mu$, for instance, $\mu = - 1.1$. For $\om = 3.5$, the 
quasienergy spectrum has been zoomed around (c) $E^F T =\pi$ and (d) $E^F T 
= 0$. They show that the modes are truly massive, and also that massless end 
modes at $E^F T = 0$ and $\pi$ can co-exist at certain values of $\mu$, 
such as $\mu = -1.1$.} \label{fig_fsl} \end{figure*}

Next, we find that as we vary $\mu$, between two successive $0$ or 
$\pi$ crossings of the quasienergy, 
we can have either non-topological phases with no end modes or
topological phases which host massless 
or massive end modes at or around $E^F T = 0$ or $\pi$ depending on 
whether $\al>1$ or $\al<1$ respectively. The divergence of $f_\al (k)$ and 
in turn of $E^F_k T$ at $k = 0$ for $\al<1$, in fact, allows the existence of 
massive Dirac edge modes even for the periodically kicked system. Furthermore, 
we observe the following from Eqs.~\eqref{eq_me} and \eqref{eq_mo}.\\

\noi (i) The $0$ and $\pi$ crossings occur at multiple values of $\mu$ as the 
number $n$ in Eqs.~\eqref{eq_me} and \eqref{eq_mo} hits different integer 
values.

\noi (ii) The multiple crossings at $E^F T = 0$ and $\pi$ result in the 
appearance of topological phases for certain ranges of $\mu$ which are 
topologically trivial in the undriven system.

\noi (iii) The gaps $\De \mu$ between $0$ or $\pi$ crossings increase 
as the frequency $\om$ increases.

\noi (iv) On decreasing $\om$, the gaps $\De \mu$ between 0 and $\pi$ 
crossings decrease, and we then find that are regions of $\mu$ where 
end modes at and around $E^F T = 0$ and $E^F T = \pi$ co-exist.\\

We now discuss the Floquet quasienergy spectrum $E^F T$ of the LRK chain with
open boundary conditions for the two cases $\al >1$ and $\al < 1$. The spectrum 
$E^F T$ is plotted versus $\mu$ for $\al = 2.1$, $V=0.2$, and 
two values of the kicking frequency, $\om = 7.0$ in Fig.~\ref{fsws1} and 
$\om =3.5$ in Fig.~\ref{fsws2}. In these figures we observe that
massless end modes are present at $E^F T= 0$ and $\pi$ for certain ranges 
of values of $\mu$. It is worth noting that the energy spectrum of the
undriven system is topologically trivial (i.e., has no MZMs) for these values 
of $\mu$, for both the frequencies. However, in contrast to the $\om = 7.0$ 
case where massless end modes do not appear at $E^F T = 0$ and $\pi$ at 
the same values of $\mu$, these two kinds of modes can co-exist for certain 
ranges of $\mu$ (for example, $\mu = -0.9$) for the case $\om = 3.5$. This 
happens due to a decrease in the gaps $\De \mu$ between crossings at $E^F T 
= 0$ and $\pi$ as $\om$ is decreased. It is interesting to 
zoom into the quasienergy spectrum near $E^F T=\pi$ (see Fig.~\ref{fsws2zt})
and $E^F T= 0$ (see Fig.~\ref{fsws2zb}), for $\om = 3.5$. These figures show 
that the modes are truly massless and also that the modes at $E^F T = 0$ and 
$\pi$ co-exist at certain values of $\mu$ such as $\mu = -0.9$.

For the case $\al < 1$, we find that the Floquet quasienergy spectrum does 
not contain any massless modes. On plotting $E^F T$ versus $\mu$ for 
$\al = 0.7$, $V=0.2$, and frequencies $\om = 7.0$ in Fig.~\ref{flsw1} and 
$\om =3.5$ in Fig.~\ref{flsw2}, we observe that massive end modes are 
present around $E^F T = 0$ and $\pi$ for certain ranges of $\mu$ where 
the undriven system is topologically trivial. However, in contrast to the 
$\om = 7.0$ case where massive end modes do not appear at $E^F T = 0$ and 
$\pi$ at the same values of $\mu$, these two kinds of modes can co-exist for 
certain ranges of $\mu$ (for example, $\mu = -1.1$) for the case $\om = 3.5$.
The gaps $\De \mu$ between crossings at $E^F T = 0$ and $\pi$ decrease as 
$\om$ is decreased. Once again, it is interesting to look more closely at the 
quasienergy spectrum near $E^F T=\pi$ (see Fig.~\ref{flw2zt}) and $E^F T = 0$ 
(see Fig.~\ref{flw2zb}), for $\om = 3.5$. The plots show that the modes are 
truly massive and not massless as in the case $\al >1$. We also observe that
that the modes at $E^F T = 0$ and $\pi$ co-exist at certain values of $\mu$ 
such as $\mu = -1.1$.

\section{Topological invariants}
\label{sec6}

The critical phase boundaries established in Sec.~\ref{sec5} for the 
periodically kicked LRK chain indicate that topologically trivial or 
non-trivial phases can lie between successive gap closings at $E^F T = 0$ 
or $\pi$ as we vary the system parameters such as $\mu$. 
To understand better whether the system lies in a topologically trivial or 
non-trivial phase, it would useful to construct bulk topological invariants 
which can characterize the different phases. Moreover, it would also be 
interesting to know if a bulk-boundary correspondence (i.e., a relation 
between the bulk topological invariants and the number of end modes) holds for
the LRK chain. To this end, we numerically determine the number and the 
(massless or massive) nature of the end modes when the system belongs to 
a topologically non-trivial phase.

We begin our analysis by calculating the bulk winding number from the 
Floquet Hamiltonian in Eq.~\eqref{eq_hamf}. The winding number is defined as
\beq \label{eq_winf}
\nu ~=~ \frac{1}{2\pi} ~\oint ~dk ~\frac{\pa_k c^y_k}{c^z_k}. \eeq
We note that the winding number defined above for the periodically kicked 
system is meaningful only for $\al>1$, and is ill-defined for $\al < 1$
due to the divergence of the polylogarithm function which appears inside
the expressions in Eq.~\eqref{eq_fspec}.
Further, calculating the winding number analytically turns out to be 
extremely difficult, and numerical approaches 
also face problems due to the divergence of the group velocity for $\al<3/2$, 
requiring extremely large system sizes for the results to convergence.

We therefore focus our attention on a second bulk invariant (see 
Ref.~\onlinecite{thakurathi}) which is known to correctly give the number 
of end modes with $E^F T = 0$ and $\pi$ for the periodically kicked 
Kitaev chain with only nearest-neighbor pairings ($\al \to \infty$). 
The momenta $k=0$ and $\pi$ play a crucial role in the definition of this
invariant since $U_k (T,0)$ can be equal to $\pm \mathbb{1}$ at only these 
two values of $k$ (here $\mathbb{1}$ denotes the $2 \times 2$ identity matrix).
Eq.~\eqref{eq_flo2x} shows that only for $\al>1$ when there is no 
divergence in the polylogarithm function, we have $U_{k=0} (T,0) = e^{i\pi b_0 
\si^z}$ and $U_{k=\pi} (T,0) = e^{i\pi b_\pi \si^z}$ since $f_{\al}(k) = 0$ 
for $k=0,\pi$. Defining $b_{0/\pi}$ in the simplest possible way, we obtain
\bea b_0 &=& \frac{2(\mu - 1)}{\om} ~+~ \frac{2V}{\pi}, \non \\
b_\pi &=& \frac{2(\mu + 1)}{\om} ~+~ \frac{2V}{\pi}, \label{b0p} \eea
where $\om = 2\pi/T$. Note that $b_\pi$ is always larger than $b_0$.
We also observe that the expressions in Eq.~\eqref{b0p} do not depend on
the value of $\al$.

We can now see that the number of integers lying between $b_0$ 
and $b_{\pi}$ for a fixed value of $\mu$, $V$ and $\om$ serves as a bulk 
topological invariant for the system with $\al>1$. It is clear that this
number is a topological invariant since it does not change under 
small deformations of the system parameters. This number can change only at 
values of $\om$ where either $b_0$ or $b_\pi$ in Eq.~\eqref{b0p} becomes equal 
to an integer. When that happens, Eq.~\eqref{eq_flo2x} becomes equal to 
$\pm \mathbb{1}$ at either $k=0$ or $\pi$. We therefore define a bulk 
topological invariant $B$ as
\beq \label{eq_ni}
B ~=~ \floor{b_{\pi}} ~-~ \floor{b_0}, \eeq
where the floor function $\floor{x}$ denotes the largest integer less than 
or equal to a real number $x$. The difference between the floor of $b_{\pi}$ 
and the floor of $b_0$ therefore counts the number of integers lying within
the range $[b_0,b_\pi]$.

\begin{figure*}[ht]
\centering
\subfigure[]{%
\includegraphics[width=.42\textwidth,height=7.5cm]{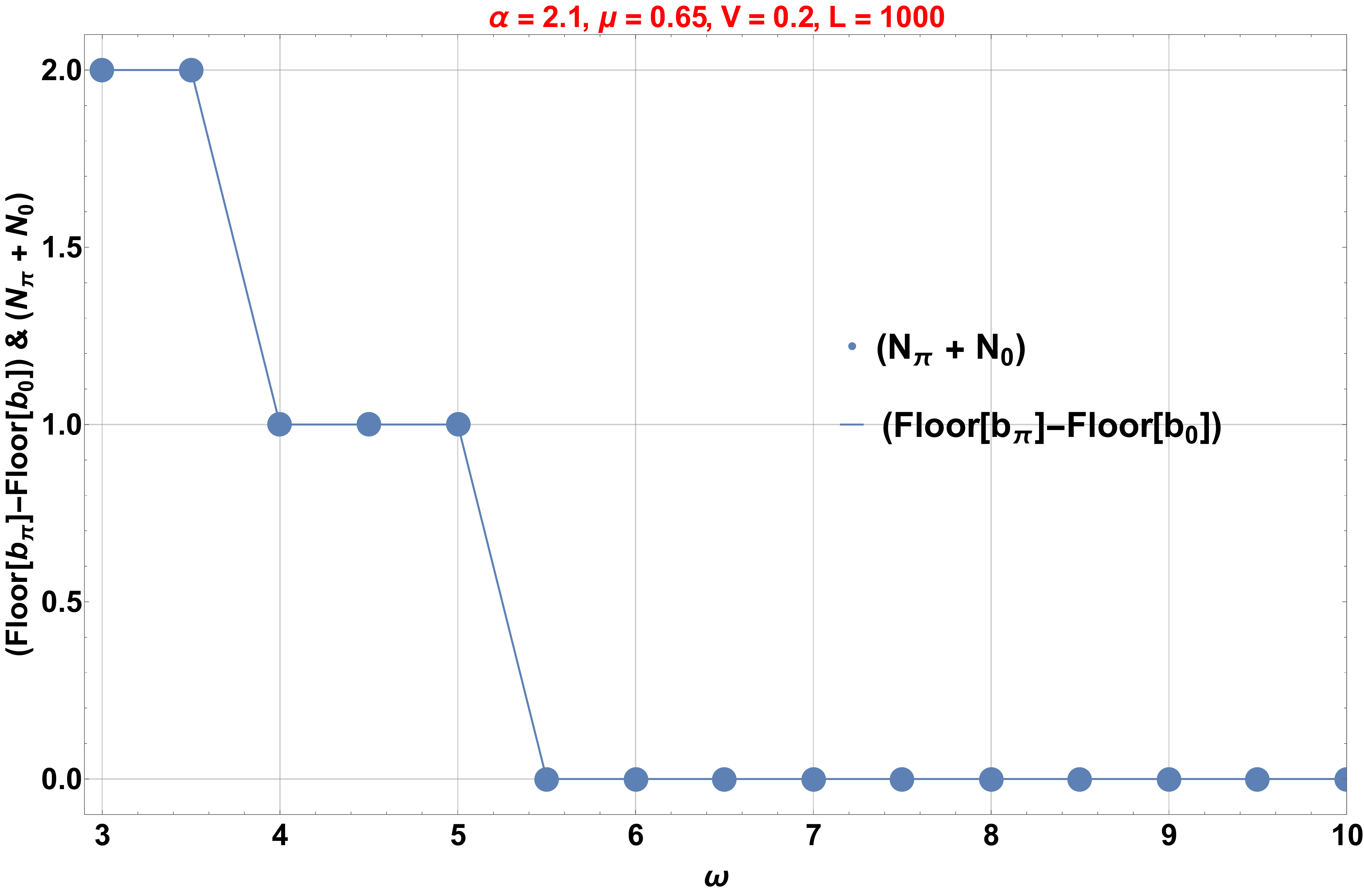}
\label{bp1}}
\hfill
\hspace*{-1cm}
\subfigure[]{%
\includegraphics[width=.42\textwidth,height=7.7cm]{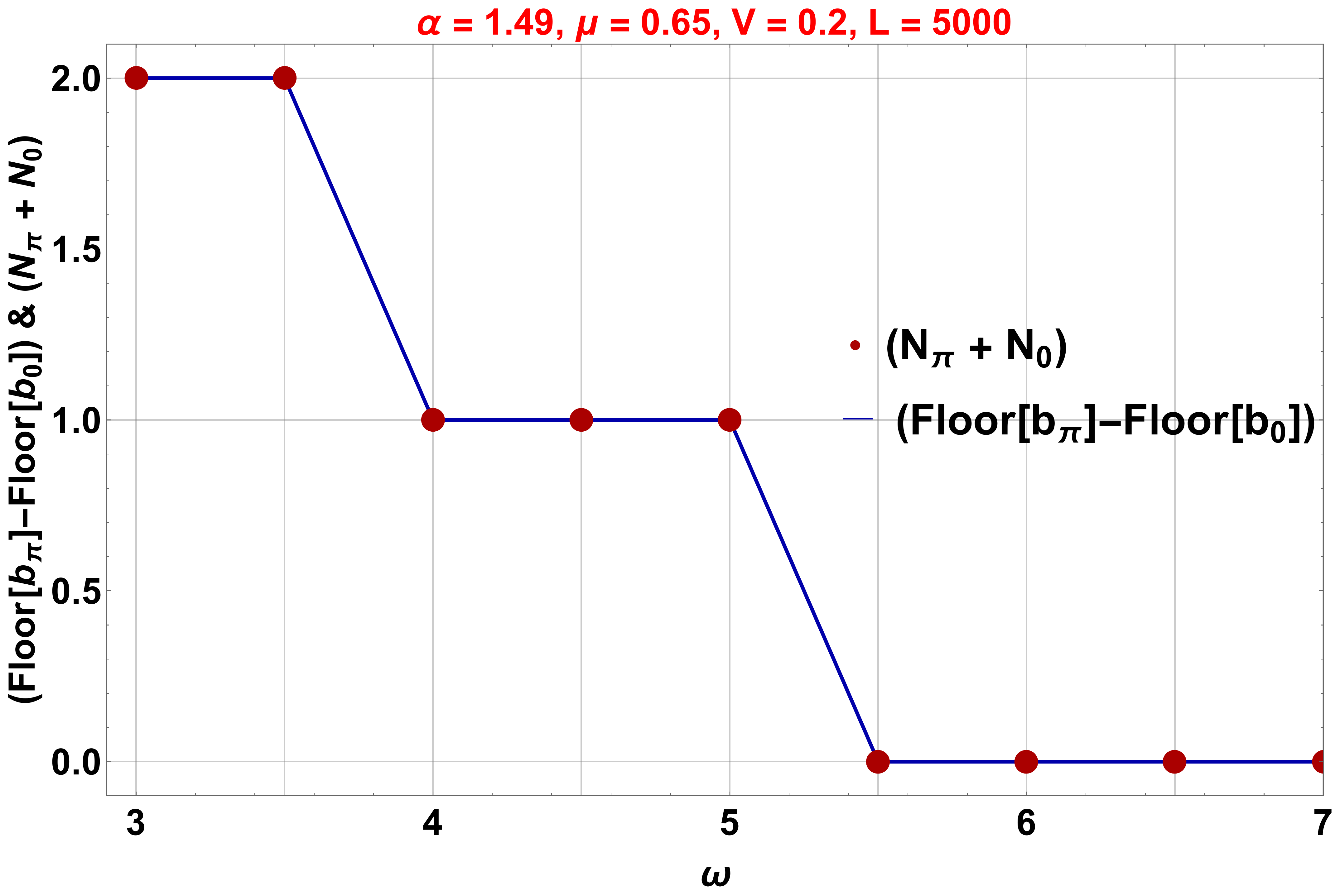}
\label{bp2}}
\caption{(a) Comparison between the analytical plot of Floor[$b_{\pi}$]-
Floor[$b_0$] (blue solid line) and the total number of massless end modes 
$N = N_0 + N_{\pi}$ (blue dots) versus $\om$, for a $5000$-site chain with 
$\mu = 0.65$, $V=2$, and (a) $\al = 2.1$ and (b) $\al = 1.49$.}
\label{BP} \end{figure*}

In Figs.~\ref{BP}, we compare the number of massless end 
modes $N = N_{\pi} + N_{0}$ (where $N_0$ and $N_\pi$ denote the numbers 
of massless end modes with $E^F T = 0$ and $\pi$ respectively at each end
an open chain and the bulk invariant $B$ as functions of $\om$, for a 
$5000$-site system with $\ga =1$, $\De = -1$, $\mu=0.65$, $V=0.2$, and (a) 
$\al = 2.1$ and (b) $\al = 1.49$. We see that the number of end modes at each 
end of the chain completely agrees with $B$ in the entire ranges of $\om$ 
shown in the two figures. We also note that in the limit $\om \to \infty$ 
(i.e., $T \to 0$), Eq.~\eqref{eq_flo2x} becomes independent of $k$; further, 
Eq.~\eqref{b0p} shows that $b_\pi \to b_0$ when $\om \to \infty$, 
hence $B=0$. This agrees with the observation that there is a maximum value 
of $\om$ beyond which there are no massless end modes in both the figures.


In Figs.~\ref{BP}, the number of Majorana end 
modes for $\om < 3$ has not been shown. As $\om$ decreases, we see from 
the expression in Eq.~\eqref{eq_ni} that the number of end modes increases. 
However, due to the presence of the long-range pairings, it becomes more and 
more difficult to identify the massless end modes as $\om$ becomes small since 
we require larger and larger system sizes to confirm if all these end modes are
really massless with Floquet eigenvalues equal to $\pm 1$ and if these 
are separated from all the other eigenvalues by finite gaps. 

\begin{figure}[ht]
\includegraphics[height=5.7cm]{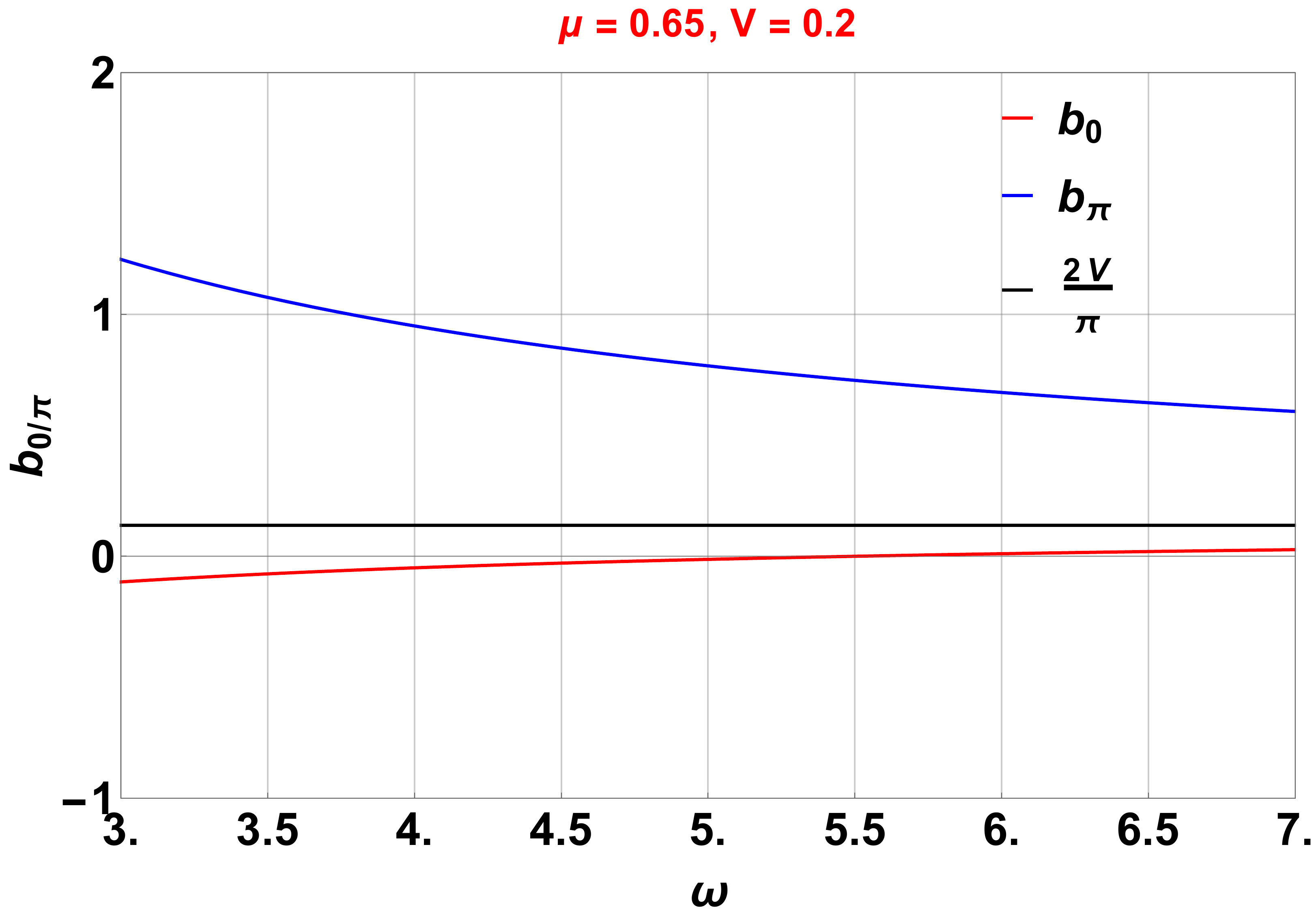}
\caption{Plots of $b_0$ (bottom red line) and $b_\pi$ (top blue curve) versus 
$\om$, for a $5000$-site chain with $\al = 1.49$, $\mu = 0.65$ and $V=0.2$. 
A horizontal black line showing $2V/\pi$ has been added as a reference.}
\label{BE} \end{figure}

There is a further refinement of the topological invariant in Eq.~\eqref{eq_ni}
which can tell us the individual values of $N_0$ and $N_\pi$, i.e., the number 
of massless modes with $E_F T = 0$ and $\pi$ respectively at each end of the 
chain. Namely, we find that the number of even and odd integers lying within 
the range $[b_0,b_\pi]$ is equal to $N_0$ and $N_\pi$ respectively; the reason
for this connection is explained in Ref.~\onlinecite{thakurathi}. This is 
illustrated in Fig.~\ref{BE} where we show
plots of $b_0$ (bottom red line) and $b_\pi$ (top blue curve) versus
$\om$, for a $5000$-site chain with $\al = 1.49$, $\mu = 0.65$ and $V=0.2$.
A horizontal black line showing $2V/\pi$ has been added as a reference. In
agreement with Eq.~\eqref{b0p}, we see that $b_0$ crosses and goes below 0 at
$\om = 5.498$, while $b_\pi$ crosses and goes above 1 at $\om = 3.782$. Hence
the number of even integers between $b_0$ and $b_\pi$ is 1 from $\om = 3$ to
$5.498$, and the number of odd integers between $b_0$ and $b_\pi$ is 1 from
$\om = 3$ to $3.782$. This is found to agree completely with the number
of massless modes $N_0$ and $N_\pi$ at each end of the LRK chain.
We have only studied values of $\om$ equal to integers and half-odd-integers,
and we find that $N_0$ is equal to 1 from $\om = 3$ to $5$ and $N_\pi$ is
equal to 1 from $\om = 3$ to $3.5$. The total number of massless end modes, 
$N_0 + N_\pi$, agrees with the results shown in Fig.~\ref{bp2}.

\section{Discussion}
\label{sec7}

In this paper, we have considered the Kitaev chain with long-range 
superconducting pairing whose strength decays as a power-law with an 
exponent $\al$. Depending upon the value of $\al$, the chain hosts massless 
Majorana or massive Dirac modes at its ends. Using exact diagonalization of 
large systems, we have established that the model is topologically similar to 
the conventional Kitaev chain with only nearest-neighbor pairing (which arises 
in the limit $\al \to \infty$) and therefore has the same topological phase 
diagram in the $\mu - \al$ plane for all $\al > 1$, not just $\al > 3/2$. For 
$\al > 1$, the system has massless Majorana modes at the two ends of the system.
For $\al < 1$, the system has a topological phase which is characterized by a 
fractional winding number and has exotic massive Dirac modes with a non-local
character. These modes arise because the long-range pairing couples the 
would-be massless modes at the two ends of the chain; this hybridization 
breaks the two-fold degeneracy at zero energy and produces massive modes 
with non-zero energies of the form $\pm E$.

Next we investigate the non-equilibrium dynamics of this system with 
long-range pairings by subjecting the chemical potential $\mu$ to periodic 
$\de$-function kicks such that the time-reversal symmetry remains intact.
We study the quasienergy spectrum as a function of $\mu$ for various values
of the system parameters and find that both the massless Majorana (for 
$\al>1$) and massive Dirac (for $\al<1$) end modes exist but only for very 
large system sizes. Interestingly, the Floquet dynamics generates new 
massless and massive end modes at quasienergy $\pi/T$ in addition to the 
modes near zero energy which are present in the absence of periodic kicks. 
Moreover, these new end modes are separated from all the other eigenvalues by 
finite gaps, and are therefore topologically robust. Furthermore, on varying 
the kicking frequency $\om$, we find that some topological phases can
emerge in which the massless or massive end modes at quasienergies $0$ and 
$\pi/T$ can co-exist. We have found the critical values of $\mu$ which
separate the non-topological and topological phases.

Motivated by the existence of bulk topological invariants (such as winding 
number) which can predict the number of massless Majorana end modes for a 
system with a time-independent Hamiltonian, we have studied if the periodically
kicked system also has topological invariants which can correctly predict the 
number of end modes, at least for $\al>1$. For $1 < \al < 3/2$, it turns out 
to be extremely difficult to compute the winding number for the periodically 
kicked system, both analytically and numerically (the latter method fails 
because the group velocity diverges if $\al<3/2$, requiring extremely large 
system sizes for the calculations to converge). Moreover, for $\al<1$, the 
winding number calculation becomes ill-defined due to the divergence of the 
polylogarithm function for the zero momentum mode. We therefore use a 
different topological invariant, first introduced in 
Ref.~\onlinecite{thakurathi}, which is easy to calculate and correctly 
predicts the number of end modes of the kicked system, provided that $\al > 1$.
We have shown that this topological invariant not only gives us the total 
number of end modes with Floquet eigenvalues equal to $+1$ and $-1$ for all 
values of the system parameters but also provides a simple condition which can 
predict the values of $\om$ at which end modes appear or disappear. In 
addition, we have found that this invariant can be written as a sum of two 
topological invariants which correctly predict the numbers of end modes with 
Floquet eigenvalues equal to $+1$ and $-1$ separately. All these topological 
invariants fail to work when $\al<1$ where the long-range pairings 
become dominant.
Finding a topological invariant which can predict the number of massive end 
modes for a periodically kicked system with $\al<1$ would thus be an
interesting and challenging problem for future study. For $\al < 1$, it 
would also be useful to understand how the energies (quasienergies) of the
massive end modes of the undriven (periodically driven) system scale
with $\mu$ close to the various phase transition points.

\vspace{.6cm}
\centerline{\bf Acknowledgments}
\vspace{.4cm}

D.S. thanks DST, India for Project No. SR/S2/JCB-44/2010 for financial support. A.D. acknowledges SERB, DST, India for financial support.

\end{document}